\documentclass[a4paper,11pt]{article}

\usepackage{fullpage}
\usepackage{setspace}
\usepackage{amsmath}
\usepackage{amsfonts}
\usepackage{hyperref}
\usepackage{graphicx}
\usepackage{import}
\usepackage{color}
\usepackage{subcaption}
\usepackage{enumitem}
\usepackage[numbers]{natbib}
\usepackage[title]{appendix}
\usepackage[affil-it]{authblk}

\newcommand{\fres}{\ensuremath{\nu_\text{r}}} % Resonant frequency.
\newcommand{\frzero}{\ensuremath{\nu_{\text{r},0}}} % Resonant frequency in limit of zero readout pwoer
\newcommand{\fread}{\ensuremath{\nu}} % Readout frequency.
\newcommand{\Pread}{\ensuremath{P_r}} % Readout power.
\newcommand{\Ures}{\ensuremath{U}} % Total energy stored in the resonator.
\newcommand{\Pt}{\ensuremath{P_t}} % Total power loss from the resonator.
\newcommand{\Pdiss}{\ensuremath{P_d}} % Power dissipated in the resonator.
\newcommand{\Qn}{\ensuremath{Q_\text{n}}} % Arbitary quality factor.
\newcommand{\Qt}{\ensuremath{Q_\text{t}}} % Total quality factor.
\newcommand{\qt}{\ensuremath{q_\text{t}}} % Total quality factor (normalised form).
\newcommand{\Qi}{\ensuremath{Q_\text{i}}} % Internal quality factor.
\newcommand{\qi}{\ensuremath{q_\text{i}}} % Normalised internal quality factor.
\newcommand{\Qc}{\ensuremath{Q_\text{c}}} % Coupling quality factor.
\newcommand{\Qtls}{\ensuremath{Q_\text{tls}}} % TLS quality factor
\newcommand{\Qtlsmin}{\ensuremath{Q_\text{tls,min}}} % Maximum internal quality factor
\newcommand{\Qqp}{\ensuremath{Q_\text{qp}}} % Quasiparticle Q
\newcommand{\Qqpth}{\ensuremath{Q_\text{qp,th}}} % Quasiparticle Q
\newcommand{\qqpth}{\ensuremath{q_\text{qp,th}}} % Quasiparticle Q / Qc
\newcommand{\Qo}{\ensuremath{Q_\text{other}}} % Qother in power law model.
\newcommand{\qo}{\ensuremath{q_\text{other}}} % Qother / Qc in power law model.
\newcommand{\Qnl}{\ensuremath{Q_\text{nl}}} % Qnl in power law model.
\newcommand{\qnl}{\ensuremath{q_\text{nl}}} % Qnl/Qc in power law model.
\newcommand{\Qimax}{\ensuremath{Q_\text{i,max}}} % Maximum internal quality factor
\newcommand{\Qtmax}{\ensuremath{Q_\text{t,max}}} % Maximum quality factor
\newcommand{\Qtmin}{\ensuremath{Q_\text{t,min}}} % Minimum quality factor
\newcommand{\Pctls}{\ensuremath{P_\text{c,tls}}} % TLS scale power
\newcommand{\Pcpl}{\ensuremath{P_\text{c,nl}}} % Scaling power for power law model.
\newcommand{\Pcqp}{\ensuremath{P_\text{c,qp}}} % Scale power for qp effects
\newcommand{\Uckin}{\ensuremath{U_\text{c,kin}}} % Scale energy for reactive non-linearity
\newcommand{\Uctls}{\ensuremath{U_\text{c,tls}}} % Scale energy for TLS
\newcommand{\Tc}{\ensuremath{T_\text{c}}} % Critical temperature
\newcommand{\nqp}{\ensuremath{n_\text{qp}}} % Quaiparticle number density
\newcommand{\nqpth}{\ensuremath{n_\text{qp,th}}} % Thermal Quaiparticle number
\newcommand{\nomega}{\ensuremath{n_\omega}} % Phonon number density
\newcommand{\nomegath}{\ensuremath{n_{\omega,\text{th}}}} % Phonon number density
\newcommand{\nstar}{\ensuremath{n_*}} % Scaling density
\newcommand{\tpb}{\ensuremath{\tau_\text{pb}}} % Pair-breaking time scale
\newcommand{\tl}{\ensuremath{\tau_\text{l}}} % Phonon lifetime
\newcommand{\Rrc}{\ensuremath{R_0}} % Recombination rate
\newcommand{\eff}{\ensuremath{\epsilon_\text{r}}} % Generation efficiency
\newcommand{\Zs}{\ensuremath{Z_\text{s}}} % Surface impedance
\newcommand{\Rs}{\ensuremath{R_\text{s}}} % Surface resistance
\newcommand{\Xs}{\ensuremath{X_\text{s}}} % Surface reactance

\newcommand{\Zref}{\ensuremath{Z_0}} % Reference impedance.
\newcommand{\Pnl}{\ensuremath{P_\text{nl}}} % Power dissipated by power law. process
\newcommand{\ytls}{\ensuremath{\xi_\text{tls}}} % y = Pr / Ptls
\newcommand{\ntau}{\ensuremath{n_\tau}} % Scaling density for limiting qp lifetime
\newcommand{\tmax}{\ensuremath{\tau_\text{max}}} % Max. qp lifetime
\newcommand{\nqpmin}{\ensuremath{n_\text{qp,min}}} % Critical number density

\begin{document}

\title{Nonlinear Effects in Superconducting Thin Film Microwave Resonators}
\author[]{C.N. Thomas, S. Withington, Z. Sun, T. Skyrme and D.J. Goldie}
\affil[]{Cavendish Laboratory, JJ Thomson Avenue, Cambridge, CB3 0HE, UK}
\affil[]{Corresponding author: \href{mailto:cnt22@cam.ac.uk}{cnt22@cam.ac.uk}}

\maketitle

\onehalfspacing

\section*{Abstract}\label{sec:abstract}

We discuss how reactive and dissipative non-linearities affect the intrinsic response of superconducting thin-film resonators.
We explain how most, if not all, of the complex phenomena commonly seen can be described by a model in which the underlying resonance is a single-pole Lorentzian, but whose centre frequency and quality factor change as external parameters, such as readout power and frequency, are varied.
What is seen during a vector-network-analyser measurement is series of samples taken from an ideal Lorentzian that is shifting and spreading as the readout frequency is changed.
According to this model, it is perfectly proper to refer to, and measure, the resonant frequency and quality factor of the underlying resonance, even though the swept-frequency curves appear highly distorted and hysteretic.
In those cases where the resonance curve is highly distorted, the specific shape of the trajectory in the Argand plane gives valuable insights into the second-order physical processes present.
We discuss the formulation and consequences of this approach in the case of non-linear kinetic inductance, two-level-system loss, quasiparticle generation, and a generic model based on a power-law form.
The generic model captures the key features of specific dissipative non-linearities, but additionally leads to insights into how general dissipative processes create characteristic forms in the Argand plane.
We provide detailed formulations in each case, and indicate how they lead to the wide variety of phenomena commonly seen in experimental data.
We also explain how the properties of the underlying resonance can be extracted from this data.
Overall, our paper provides a self-contained compendium of behaviour that will help practitioners interpret and determine important parameters from distorted swept-frequency measurements.

\section{Introduction}\label{sec:intro}

Superconducting thin-film microwave resonators are being developed for a wide range of applications.
For example, in astronomy, large arrays of Kinetic Inductance Detectors (KIDs) are being developed for ultra-low-noise measurements (100-800 GHz) of the polarisation state of the cosmic microwave background radiation\,\cite{galitzki2014next,steinbach2018thermal,tang2018fabrication}, to carry out galaxy surveys in the sub-millimetre-wave region\,\cite{monfardini2010nika,calvo2016nika2,endo2019wideband,brien2018muscat}, and for energy and time resolved optical and x-ray photon counting experiments in high energy astrophysics\,\cite{mazin2013arcons,ulbricht2015highly}.
Arrays of superconducting resonators coupled to Superconducting Quantum Interference Devices (SQUIDs) provide a convenient way of reading out large arrays of ultra-low-noise devices that are not themselves easily multiplexed, such as Transition Edge Sensors\,\cite{mates2017simultaneous,bruijn2018lc,dobbs2012frequency}.
In quantum computing, superconducting resonators are being coupled to tunnel junctions to create qubits \,\cite{martinis2009superconducting}, and to embedded spin systems to create memory elements\,\cite{morton2018storing}.
More generally, thin-film superconducting resonators are a natural system for exploring chip-based Quantum Electrodynamics (QED)\,\cite{wallraff2004strong,goppl2008coplanar}, and are being realised in exotic combinations, such as superconducting electromagnetic resonators coupled to micromechanical cantilevers for studying quantum-statistical processes\,\cite{regal2008measuring}.

Not only are the applications varied, the physical realisations are diverse.
Superconducting resonators can take the form of microstrip or coplanar transmission lines, shaped conductors in waveguide cavities, or even lumped-element components based on thin-film inductors and capacitors.
The metallic films usually take the form of Ultra High Vacuum (UHV) deposited normal metals or superconductors (Nb, Al, Ta, Ti, NbN, NbTiN) laid down on dielectric substrates (Si, SiN, and Sapphire).
The conductors can also take the form of proximitised multilayers (TiAl, TiAu, MoAu) for customising the bulk properties of films, and the substrates can be irradiated (nitrogen-vacancy centres in diamond) or surface implanted with dopants (P,Bi) to enable spin-system coupling\,\cite{morton2018storing}.

A crucial point is that when superconducting resonators are measured, they often do not behave in a simple linear way having a near-perfect Lorentzian response, but instead show transmission and reflection coefficients that display peculiar shapes in the complex plane.
Moreover, their behaviour changes as the readout power is increased, and often the resonance curves switch hysteretically between two stable states as the readout frequency is swept up and down.
These effects can vary significantly between two notionally identical devices, emphasising the importance of fabrication methods and conditions.
It follows that although a device may be designed on the basis of near-ideal behaviour, the actual behaviour is influenced strongly by the non-ideal characteristics of the materials used.
Understanding  these `second order' effects is an essential feature of any development programme, particularly when quantum-limited operation is sought.

In this paper, we review the theoretical description of superconducting resonators, and show how a simple model based on the notion of power and energy dependent resonance frequency and quality factor can account for a wide variety of phenomena seen.
We show that a considerable amount of physical information is contained in the behaviour of the quality factor, not just in the resonant frequency, as external parameters, such as the readout power, are changed.
In fact, particular shapes in the complex plane are characteristic of different physical mechanisms, and it is highly desirable to be able to identify these easily when carrying out experiments, or when, say, characterising films and geometries.
We describe a range of methods for extracting physical information from distorted resonance curves, which can then be used for optimising performance, and for predicting operational aspects of behaviour such as optimal readout power, small signal nonlinearity, and noise.

\section{Preliminaries}\label{sec:preliminaries}

\subsection{Definitions of key symbols used throughout the paper}\label{sec:definition_of_symbols}

\begin{description}[font=\normalfont, leftmargin=1.5cm, style=nextline]
\item[$\fres$]
Resonant frequency.
\item[$\frzero$]
Resonant frequency in the limit of zero readout power.
\item[$\fread$]
Measurement/readout frequency.
\item[$\Pread$]
Applied power at readout port.
\item[$\Ures$]
Total energy stored in the resonator.
\item[$\Pt$]
Total power loss from the resonator.
\item[$\Pdiss$]
Power dissipated internally in the resonator.
The difference between $\Pt$ and $\Pdiss$ is the power loss to the readout circuit.
\item[$\Qn$]
General notation for quality factor contribution from a particular loss mechanism.
\item[$q_n$]
Normalised quality factor $q_n = Q_n / \Qc$, where $\Qc$ is the coupling quality factor.
\item[$\Qt$]
Total resonator quality factor, accounting for all losses.
\item[$\Qi$]
Internal (or intrinsic) quality factor resulting from all internal losses ($\Pdiss$).
\item[$\Qc$]
Coupling quality factor associated with power loss to the readout circuit.
\item[$\Qtls$]
Quality factor from two-level-system (TLS) loss.
\item[$\Qtlsmin$]
Value of $\Qtls$ in the limit of zero readout power.
\item[$\Qqp$]
Quality factor from losses in the quasiparticle systems of the superconductors in the resonator.
This is associated with the Ohmic losses on any superconducting surfaces.
\item[$\Qqpth$]
Value of $\Qqp$ when the quasiparticle population is purely thermal.
\item[$\Qnl$]
Used to indicate the quality factor contribution from the model non-linearity of Section \ref{sec:power_law_models}.
\item[$\Qo$]
Used throughout to indicate the contribution from internal loss mechanisms other than the particular mechanism of interest.
Assumed independent of applied readout power.
\item[$Q_\text{n,min}$, $Q_\text{n,max}$]
Minimum and maximum value of $\Qn$ as a function of readout power.
\item[$P_\text{c,n}$]
Scale power for a nonlinear effects due to a particular mechanism.
In this paper $n=\text{tls}$, $\text{qp}$ and $\text{nl}$, corresponding to quasiparticle losses, TLS losses and losses due to the power-law model process.
\item[$U_\text{c,n}$]
Scale energy for a nonlinear effects due to a particular mechanism.
\item[$x_0$]
`Applied' fractional detuning, as defined by $(\nu - \frzero) / \frzero$.
\item[$x$]
`Realised' fractional detuning, as defined by $(\nu - \fres) / \fres$.
Here $\fres$ is the instantaneous value of the resonant frequency.
Because $\fres$ can vary depending on the energy stored or dissipated in the resonator, $x$ and $x_0$ are only equal in the absence of reactive nonlinearities or in the limit of zero readout power.
It is the realised fraction detuning that determines the measured $S$ parameters.
\item[$y$]
Defined as $y = \Qc x$.
Scaling $x$ by $\Qt$ yields the realised detuning as measured in linewidths from the resonator: when $\Qt x = 1$ the readout frequency is tuned a resonance-width above the centre frequency.
Since we assume $\Qc$ is fixed and by definition $\Qt \leq \Qc$, $y$ specifically corresponds to the \emph{maximum} value $\Qt x$ can take for all readout powers.
\item[$y_0$]
Defined as $y_0 = \Qc x_0$.
The applied detuning measured in linewidths.
\item[$\Tc$]
Superconducting critical temperature.
\item[$\Delta$]
Superconducting gap energy.
\item[$\nqp$]
Quasiparticle number density in the active volume of the device.
\item[$\nqpth$]
$\nqpth$ is the value of $\nqp$ in the limit of zero readout power, i.e. arising from thermal processes alone.
\item[$\nstar$]
Value of $\nqpth$ at which $\Qqp = \Qc$.
\item[$\nomega$]
Number density of pair-breaking phonons with energy in excess of twice the superconducting gap energy, $2\Delta$, in the active volume of the resonator.
\item[$\nomegath$]
$\nomegath$ is the value of $\nomega$ in the limit of zero readout power, i.e. arising from thermal processes alone.
\item[$V$]
Volume of the active region of the device.
\item[$\tpb$, $\tl$, $\Rrc$, $\eff$]
Parameters in the Rothwarf-Taylor model.
$\tpb$ is the pair-breaking lifetime, $\tl$ is the timescale on which pair-breaking phonons are lost to scattering, $\Rrc$ is the quasiparticle recombination rate and $\eff$ is the efficiency with which dissipated readout power is converted to pair-breaking phonons.
\item[$\Zs = \Rs + i \Xs$]
Surface impedance of the superconductor.
$\Rs$ and $\Xs$ are the resistive and reactive components, respectively.
\end{description}

\subsection{Quality factor}\label{sec:qfactors}

Quality factor is a well known measure of energy loss in resonant circuits.
When the loss is due to a combination of dissipative processes, it is common to define a $Q$-like measure for each of the processes.
However, a range of conventions exist, and so in this section we outline the terminology that will be used in this paper.

Let $U$ be the energy stored in a resonator having resonant frequency $\fres$.
If the average total power dissipated is $\Pt$, then the overall quality factor $\Qt$ is defined by
\begin{equation}\label{eqn:def_qt}
	\Qt = \frac{2 \pi \fres U}{\Pt}.
\end{equation}
For a resonator coupled to an external circuit, $\Pt$ includes the energy lost to that circuit.
Now assume that the total loss is due to a number of different dissipative processes, such that $\Pt = \sum_n P_n$.
Then
\begin{equation}\label{eqn:qt_as_reciprocal_sum}
	\Qt^{-1} = \sum_n Q_n^{-1},
\end{equation}
where
\begin{equation}\label{eqn:def_qn}
	Q_n = \frac{2 \pi \fres U}{P_n}
\end{equation}
are effective quality factors, or equivalently the actual quality factor when only the $n$'th loss is present.
The total \emph{internal} quality factor, $\Qi$ characterises losses `internal' to the resonator, in the sense they would still exist if the resonator were isolated from the readout circuit.
$\Qi$ may comprise contributions from several microscopic processes: Ohmic loss and dielectric loss are examples. $\Qi$ is also commonly referred to as the \emph{unloaded}\,\cite{pozar2012microwave} or \emph{intrinsic}\,\cite{vanier2011intrinsic} quality factor.
The total \emph{coupling} quality factor, $\Qc$, is associated with the power lost from the resonator to the readout circuit.
This loss is a pure feature of coupling and exists by virtue of reciprocity -- if energy can be transferred into the resonator, it can also be transferred out of the resonator.
$\Qc$ may also comprise loss by several mechanisms, e.g. to different ports of a multiport readout system.
Based on these definitions, we can always make the division
\begin{equation}\label{eqn:qt_in_terms_of_qi_and_qc}
	\Qt^{-1} = \Qi^{-1} + \Qc^{-1}.
\end{equation}
$\Qt$ in this instance is also sometimes referred to as the \emph{loaded}-Q of the device\,\cite{pozar2012microwave}.
A device is said to be undercoupled or overcoupled  if $\Qc > \Qi$ or $\Qc < \Qi$, respectively.
Throughout this paper, we will use lower-case $q$ to denote a quality factor normalised to the coupling quality factor:
\begin{equation}\label{eqn:def_normalised_q}
	q_n = \frac{Q_n}{\Qc}.
\end{equation}
$q_n$ is a measure of the degree to which power lost through mechanism $n$ compares with the power lost to the readout circuit.

\subsection{Microwave scattering parameters of common resonator circuits}\label{sec:sparameters}

Consider a device comprising a resonator embedded in, and lightly coupled to, a lossless, reciprocal, multiport readout circuit.
Temporal coupled mode theory\,\cite{haus1984waves} can be used to show that the microwave scattering parameters $\{S_{mn}\}$ at the external ports of the overall circuit have the general form
\begin{equation}\label{eqn:s_matrix_general}
	S_{mn} (\nu) = \Gamma_{mn} + \frac{K_{mn}}{1 + 2 i \Qt x}.
\end{equation}
$\Gamma$ is the scattering matrix of the isolated readout circuit, $K$ is a symmetric coupling matrix, and $x$ the realised fractional detuning,
\begin{equation}\label{eqn:def_detuning_x}
x = \frac{\nu - \fres}{\fres},
\end{equation}
where $\nu$ is the readout frequency.
We will also refer to the realised detuning $y$ in coupling-Q linewidths, which we define by
\begin{equation}\label{eqn:def_y}
	y = \Qc x.
\end{equation}
Strictly we are making a single-pole approximation by neglecting the contribution from the pole at $\nu = -\fres$, requiring $\Qt \gg 1$.

(\ref{eqn:s_matrix_general}) describes a very wide range of devices, but for illustrative purposes we will use the specific example of an embedding circuit having 2 external ports.

\begin{figure}
\centering
\includegraphics[width=8cm]{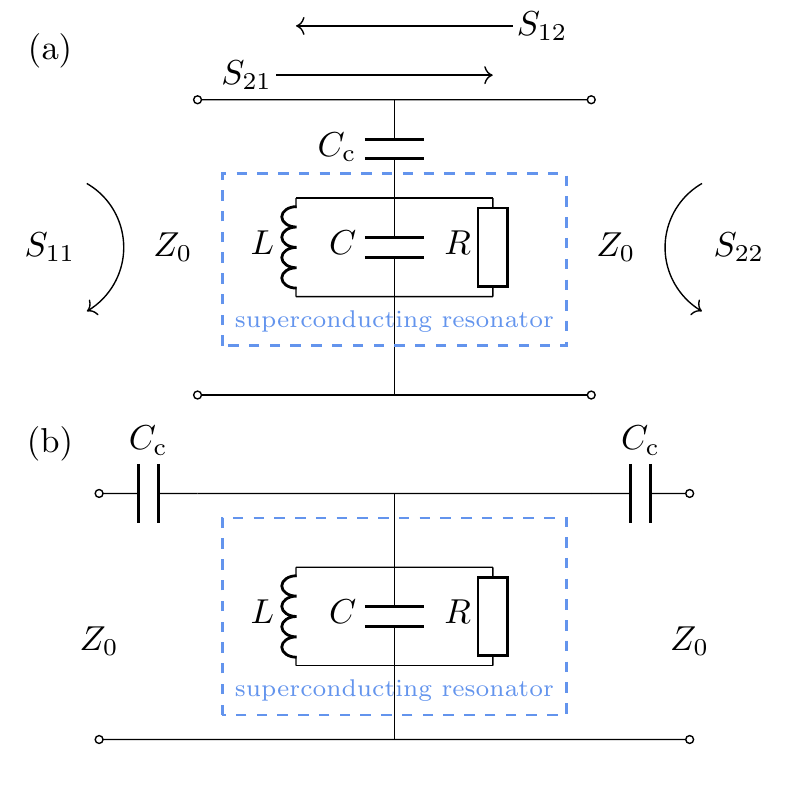}
\caption{\label{fig:different_resonator_geometries}
(a) the LCR tank represents the superconducting resonator, which could be a shorted quarter-wave superconducting transmission line\,\cite{gao2008equivalence,endo2019wideband}, an open-ended half-wave line\,\cite{endo2013chip,de2014evidence} or an implementation in discrete components\,\cite{marsden2012optical,hornsby2018initial}.
The LCR tank is lightly capacitively shunt-coupled across the readout transmission line, giving a null in transmission ($S_{21}$ and $S_{12}$) on resonance.
(b) the LCR tank circuit represents a superconducting half-wave line that is open at both ends.
This is lightly capacitively coupled in series in the readout line, giving a maximum in transmission on resonance.
Inductively coupled implementations of both designs are also possible\,\cite{doyle2008lumped,bothner2013inductively}.
}
\end{figure}

In the case of a short-circuited $\lambda/4$ `resonator', with a series coupling capacitor, connected in parallel with a through transmission line, the equivalent circuit takes the form of Figure \ref{fig:different_resonator_geometries} (a), and the scattering elements of the whole device become
\begin{equation}\label{eqn:mkid_s11}
	S_{11} = S_{22} = - \frac{\Qt}{\Qc} \frac{1}{1 + 2 i \Qt x}
\end{equation}
and
\begin{equation}\label{eqn:mkid_s21}
	S_{12} = S_{21} = 1 + S_{11}
	= 1- \frac{\Qt}{\Qc} \frac{1}{1 + 2 i \Qt x},
\end{equation}
which displays a maximum in reflection $S_{11} = S_{22} = - \Qt / \Qc $ and a minimum in transmission $S_{12} = S_{21} = 1- \Qt / \Qc$ at resonance: remembering that $\Qt \leq \Qc$.
An optimally coupled resonator $\Qt = \Qc$ displays near ideal behaviour, reducing the transmitted signal to zero at resonance.
This is a good model of many devices, such as kinetic inductance detectors (KIDs), independent of the specific physical realisation\,\cite{zmuidzinas2012superconducting}.

In the case of a $\lambda/2$ `resonator', with two series coupling capacitors, connected in series with a through transmission line, the equivalent circuit takes the form of Figure \ref{fig:different_resonator_geometries} (b), and the scattering elements of the whole device become
\begin{equation}\label{eqn:mkid_s11_s}
    S_{11} = S_{22} = 1- \frac{\Qt}{\Qc} \frac{1}{1 + 2 i \Qt x},
\end{equation}
and
\begin{equation}\label{eqn:mkid_s21_s}
    S_{12} = S_{21} = - \frac{\Qt}{\Qc} \frac{1}{1 + 2 i \Qt x},
\end{equation}
which displays a  minimum in reflection $S_{11} = S_{22} = 1 - \Qt / \Qc $ and maximum in transmission $S_{12} = - \Qt / \Qc$ at resonance, illustrating the duality of parallel- and series-resonant circuits.

In many devices, one seeks a resonant notch that approaches zero, or a resonant peak that approaches unity, and in both of these cases, the coupling quality factor must be chosen to dominate the losses, which limits the operating $\Qt$ to a value lower than that implied by $\Qi$.

\subsection{Non-linear behaviour}\label{sec:nl_behaviour}

Non-linear behaviour manifests itself as variations in the $\{S_{mn}\}$ as the amplitude of the readout signal is changed.
For example, swept-frequency measurements of complex-valued scattering parameters with a vector network analyzer (VNA) can lead to traces that vary with readout power.
Numerous distorted and hysteretic resonance shapes can occur \cite{swenson2013operation,de2010readout,thompson2013dynamical}.
Here we explain many of the observed effects, and in particular consider the broad category of nonlinear behaviour that can be described as a dependence of the resonance frequency and/or quality factor on the power dissipated $\Pdiss$ internally (as distinct from the total power flowing out of the resonator, $\Pt$, which also includes the coupling loss): $\fres(\Pdiss)$  and $Q_n(\Pdiss)$ respectively.
It is clear that the dissipated power can be calculated once the scattering parameters are known.

In some cases, such as heating, the dependence on $\Pdiss$ is direct.
However, it follows from (\ref{eqn:def_qt})--(\ref{eqn:def_qn}) that $U$ and $\{P_n\}$ can all be expressed in terms of $\Pdiss$ provided the $\{\Qn\}$ are known, and so the resonant frequency and quality factor can be written in terms of $\Pdiss$ even for mechanisms that do not involve heating directly.
We will refer to changes in resonant frequency with dissipated power,  $\fres(\Pdiss)$, as \emph{reactive non-linearities}, as they are primarily caused by changes in the reactive elements of a resonator.
This will be illustrated for specific cases later.
Equivalently, we will refer to changes in quality factor with dissipated power, $\Qn (\Pdiss)$, as \emph{dissipative non-linearities}, as they are primarily caused by changes in the resistive elements of a resonator.
In this context, we will make two assumptions: (i) The coupling quality factor exhibits no non-linear behaviour, which is true for most devices because the coupling is via a near-perfect capacitance, self-inductance, or mutual inductance.
Modifying the forthcoming analysis to relax this assumption is not in itself difficult, but adds a significant algebraic overhead that distracts from the main results.
(ii) The scattering parameters are described by the functional form given in (\ref{eqn:s_matrix_general}), but with nonlinearity occurring through $\fres(\Pdiss)$  and $\Qn (\Pdiss)$ under all conditions.
Physically, this corresponds to the situation where the circuit topology remains constant, and it is only the component values that change with readout signal level.
Within this framework, the values of $\{ S_{mn} \}$ can be found for a given applied signal level through finding self consistent solutions to (\ref{eqn:s_matrix_general}), and $\fres(\Pdiss)$  and $\Qn (\Pdiss)$.
Indeed it is this generic mechanism that creates, under different conditions, many of the physical phenomena seen.

\subsection{Non-linearity in the Argand plane}\label{sec:nl_in_argand_plane}

A characteristic of linear resonant circuits is that the scattering parameters all trace out circular paths in the Argand plane as a function of frequency: only the centres and radii change with the circuit topology and circuit parameters.
This behaviour occurs because expressions having the form of (\ref{eqn:s_matrix_general}) constitute bilinear maps.

\begin{figure}
\centering
\includegraphics[width=\textwidth]{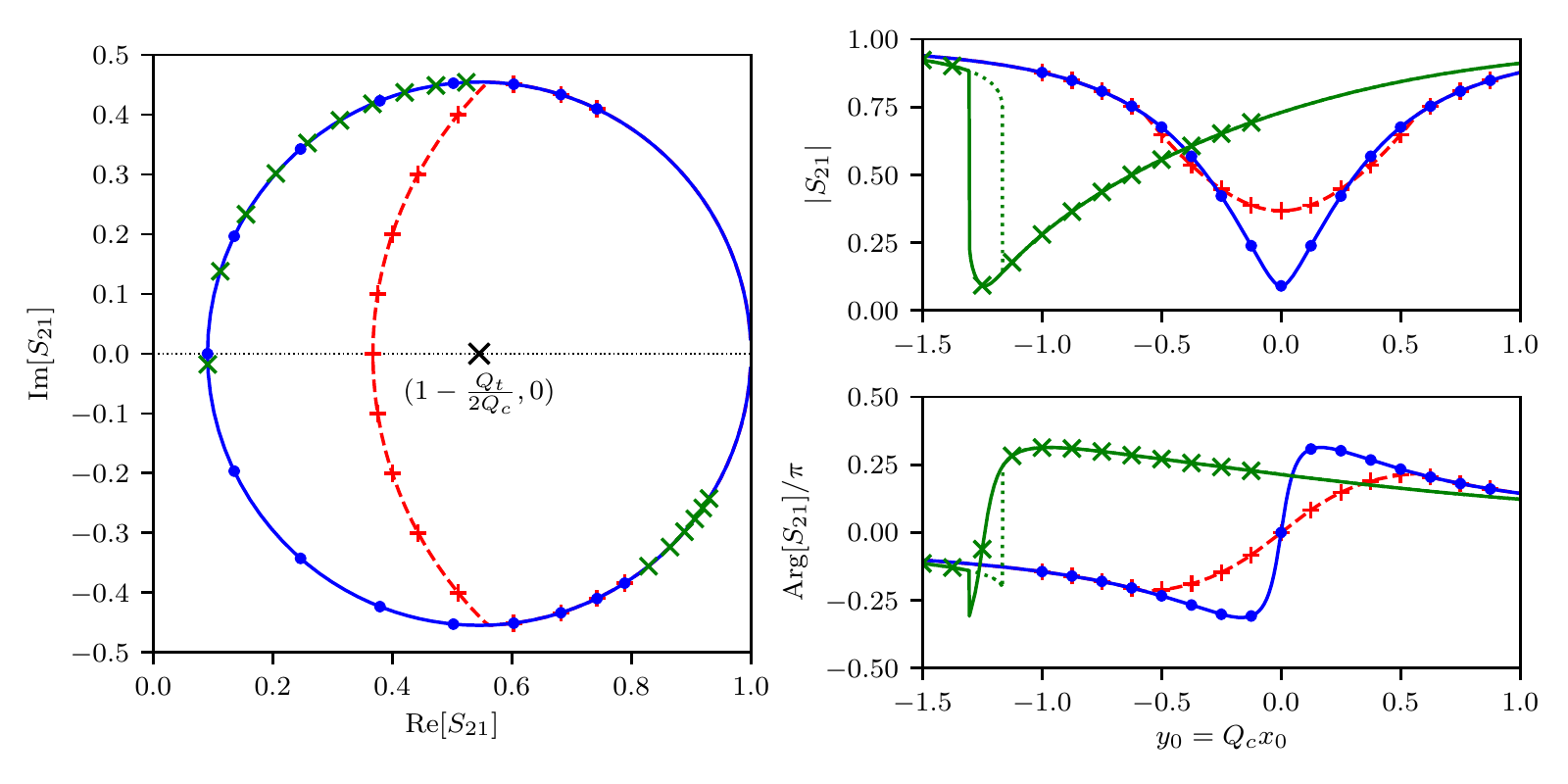}
\caption{\label{fig:resonance_circle_distortion}
$S_{21}$ as a function of frequency in the Argand plane on the left, and in the form $|S_{21}|$ and $\text{Arg}[S_{21}]$ versus $y_0$ on the right. $y_0$ is the applied detuning in (minimum) linewidths relative to the resonant frequency at infinitesimal readout power.
Blue (solid) lines show the ideal linear behaviour, with the circles indicating a set of evenly space frequency points.
Green lines on the right show the behaviour when the non-linearity is purely reactive, with the diagonal crosses indicating a set of evenly spaced frequency points (same points indicated on the left).
Solid lines show the curve measured sweeping down in frequency, while the dashed lines show the curve on sweeping up.
Red (dashed) line shows the behaviour for a hypothetical, purely dissipative, non-linearity, with the horizontal crosses indicating a set of evenly spaced frequency points.
Note that $S_{21}$ traces clockwise with increasing frequency.
}
\end{figure}
%}

To illustrate this feature consider $S_{21}$ for a parallel resonant circuit in the linear regime, as shown by the blue (solid) lines in Figure \ref{fig:resonance_circle_distortion}.
The left diagram shows the data in the Argand plane, while the right diagram shows the equivalent plots of transmission magnitude and phase as a function of the detuning in linewidths, $y_0$, relative to the resonant frequency with infinitesimal readout power.
From (\ref{eqn:mkid_s21}) we can derive
\begin{equation}\label{eqn:circle_constraint}
	|S_{21} - C | = \frac{\Qt}{2\Qc}
\end{equation}
and
\begin{equation}\label{eqn:circle_angle}
	\theta = \text{Arg} [S_{21} - C]
	= -\tan^{-1} \left( \frac{4 \Qt x}{1 + \{ 2 \Qt x \}^2} \right)
\end{equation}
where
\begin{equation}\label{eqn:cricle_centre}
	C = 1 - \frac{\Qt}{2\Qc}.
\end{equation}
(\ref{eqn:circle_constraint}) implies $S_{21}$ is constrained to lie on a circle, with $C$ the centre.
$\theta$, as defined, is the angle subtended by $S_{21}$ at $C$ as measured anticlockwise from the real axis; (\ref{eqn:circle_angle}) therefore describes the motion of $S_{21}$ around the circle as a function of frequency.
The blue circles in Figure \ref{fig:resonance_circle_distortion} indicate the value of $S_{21}$ at a set of evenly spaced frequency points spanning the resonance with $S_{21}$ moving clockwise around the circle as a function of frequency.

Non-linear behaviour can result in the resonance circle becoming distorted.
First we note that for purely reactive nonlinear behaviour, with $\Qt$ invariant over a sweep, (\ref{eqn:circle_constraint}) still constrains $S_{21}$ to lie on a circle.
The motion of $S_{21}$ around the circle with frequency may change, with the green diagonal crosses in Figure \ref{fig:resonance_circle_distortion} indicating, for example, how the frequency points corresponding to the blue circles might move.
Hysteresis with sweep direction may also be observed, and some points of the circle may even become inaccessible\,\cite{swenson2013operation}.
The radius and centre of the circle contain important information, even though the resonance curve is hysteretic.
(\ref{eqn:circle_constraint}) shows that only nonlinear dissipative behaviour can distort $S_{21}$ from a circular path.
For example, the red (dashed) lines in Figure \ref{fig:resonance_circle_distortion} show hypothetical curves for a device where $\Qt$ decreases with dissipated power, causing the effective radius of the `circle' to decrease closer to resonance.
In fact, two characteristic circles seem to be present.
In addition, (\ref{eqn:circle_angle}) indicates that dissipative non-linearities can also influence the rate at which $S_{21}$ moves around the circle in the same way as reactive non-linearities.
The preceding discussion applies equally well to any scattering parameter of any device described by (\ref{eqn:s_matrix_general}).
In what follows we will show that different dissipative processes produce characteristic distortions, making the shapes, radii, and centres of resonance `circles' powerful diagnostics of underlying physical mechanisms.

\section{Distortions in swept-frequency S-parameter measurements}\label{sec:distortion}

\subsection{Origin of distortion}\label{sec:origin}

\begin{figure}
\centering
\includegraphics[width=\textwidth]{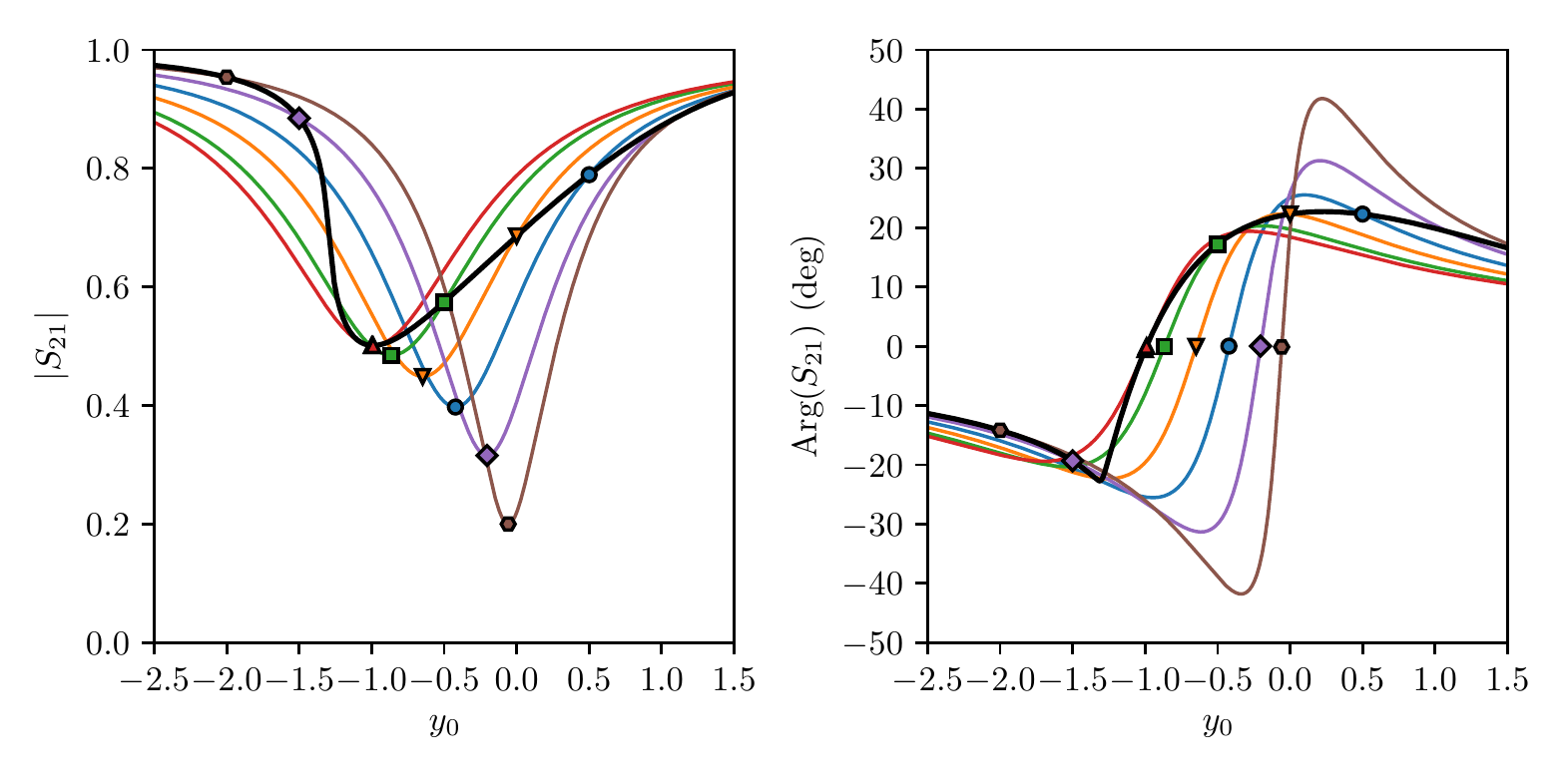}
\caption{\label{fig:origin_of_distortion}
The black lines in each plot shows the measured, distorted, resonance trace.
Each of the coloured curves indicate the behaviour of the `underlying' single-pole resonance as the trace is swept out.
Each of these curves is calculated assuming constant quality factor and resonant frequency equal to $\Qi(\Pdiss)$ and $\fres(\Pdiss)$ of the resonator at the points indicated by the markers on the black lines.
By definition, each coloured curve intersects the black line at the location of the corresponding marker.
The second set of markers, lying purely on the coloured lines, have been included simply to allow each underlying curve to be matched to the corresponding intersection point more easily.
}
\end{figure}

Consider an idealised model of a swept-frequency S-parameter measurement with a VNA or homodyne readout system \,\cite{zmuidzinas2012superconducting}.
The device under test is a two-port non-linear resonator of the type described in Section \ref{sec:qfactors}, with generalised S-parameters given by (\ref{eqn:mkid_s11}) and (\ref{eqn:mkid_s21}).
Assume that all S-parameters and power-wave amplitudes are defined relative to reference impedance $\Zref$.

A sinusoidal voltage source with frequency $\nu$ and real output impedance $\Zref$ is used to drive the resonator at port 1 and a load of impedance $\Zref$ is connected to port 2.
Under these conditions, $S_{11} = b_1 / a_1$ and $S_{21} = b_2 / a_1$ are the scattering parameters referenced to $\Zref$, and $a_1$, $b_1$ and $b_2$ are the measured complex amplitudes of the incident travelling wave at port 1, outgoing wave at port 1 and outgoing wave at port 2, respectively.
Assume that the source frequency is swept to measure $S_{11}(\nu)$ and $S_{21}(\nu)$ while keeping the readout power $\Pread = |a_1|^2$ constant.

If the resonator is driven into a non-linear regime, the variation in the dissipated power with frequency will generally result in distortion of the measured data compared with (\ref{eqn:mkid_s11}) and (\ref{eqn:mkid_s21}).
Visually, we will record resonance curves that look like the red and green lines on the right in Figure \ref{fig:resonance_circle_distortion}, rather than the blue line.
Now consider the mechanism by which this distortion arises in our framework.

The dissipated power is the difference between the outgoing power at ports 1 and 2 and the incoming power at port 1,
\begin{equation}\label{eqn:pdiss_full}
	\Pdiss = |a_1|^2 - |b_1|^2 - |b_2|^2
	= (1 - |S_{11}|^2 - |S_{21}|^2) \Pread.
\end{equation}
Using (\ref{eqn:mkid_s11}) and (\ref{eqn:mkid_s21}) to substitute for the S-parameters, we obtain
\begin{equation}\label{eqn:p_res}
	p = \frac{2 \eta}{(1 + \eta)^2 + 4 y^2},
\end{equation}
where $p = \Pdiss / \Pread$ is the normalised power dissipation, $\eta = q_i^{-1} = \Qc / \Qi$ is the normalised internal dissipation factor and $y$ is the detuning in linewidths as defined in Section \ref{sec:qfactors}.
(\ref{eqn:p_res}) indicates that the dissipated power peaks sharply at $2 \eta / (1 + \eta)^2$ as the source frequency is tuned through resonance, and falls to zero either side.
In the same notation
\begin{equation}\label{eqn:s11_and_s21_p_notation}
	S_{11} = S_{21} - 1 = -\frac{1}{1 + \eta + 2 i y}.
\end{equation}
However,  $\eta$ and $y$ are both functions of the dissipated power through their dependence on $\Qi$ and the resonant frequency, with $y$ also dependent on the readout frequency.
Since the incident readout power is fixed in a frequency sweep, we can alternatively express this as a dependence on normalised dissipated power: $\eta(p)$ and $y(\nu, p)$, respectively.
When the source frequency is changed to a new value, the dissipated power (and with it $\Qi$ and the resonant frequency) evolves to a new equilibrium.
It follows from (\ref{eqn:p_res}) that the normalised dissipated power in the final state, $p_0$, must satisfy the condition
\begin{equation}\label{eqn:p_steady_state}
	p_0 = \frac{2 \eta}{(1 + \eta(p_0))^2 + 4 y(\nu, p_0)^2},
\end{equation}
at the readout frequency $\nu$.
The dynamical process by which the circuit moves to the equilibrium condition depends on the physical realisation, and an example has been described by Thompson\,\cite{thompson2013dynamical}.
In the subsequent discussion we will assume that $\nu$ is always swept slowly enough that (\ref{eqn:p_steady_state}) is satisfied at all points, for example that there are no thermal delays, and we will use the notation $p$, rather than $p_{0}$, without confusion.
Note that there may be multiple solutions of (\ref{eqn:p_steady_state}), in which case hysteretic behaviour can occur.

The proceeding discussion indicates how quality factor and resonant frequency can become functions of the measurement frequency, giving distorted resonance curves of the kind shown in Figure \ref{fig:resonance_circle_distortion}.
There is a simple visualisation of the process: Figure \ref{fig:origin_of_distortion}.
At each measurement frequency, the circuit has a simple Lorentzian resonance, and the measurement simply samples one point on this resonance.
If the measurement frequency is changed, the underlying resonance curve changes, giving rise to a new sample taken from a new Lorentzian.
Thus the observed shape is merely a manisfistation of the fact that a simple underlying Lorentzian is sweeping through the sample points taken: the underlying curve being swept out, as defined by (\ref{eqn:mkid_s11}) and (\ref{eqn:mkid_s21}), changes as we proceed through the swept-frequency measurement process.
Crucially, the origins of the distortions lie in translations and rescalings of the underlying linear resonance, and this puts constraints on the observed behaviour.
In fact, certain features of the linear resonant behaviour carry over to even highly distorted curves, as we will now show.

This model emphasises why the experimenter does not usually have direct control over the detuning $x$ as given by (\ref{eqn:def_detuning_x}): they can set $\nu$, but in the presence of reactive non-linearities they may not know $\fres$.
We will refer to $x$ throughout as the `realised' detuning at a particular frequency.
It is $x$ that is used in (\ref{eqn:mkid_s11})--(\ref{eqn:mkid_s21_s}) to calculate $S$, and which determines the underlying resonance curve at a point as illustrated in Figure \ref{fig:origin_of_distortion}.
However, it is still often useful to express a readout frequency as a detuning.
To do so we can use the limiting value $\frzero$ of the readout frequency at zero (or sufficiently low) readout power as our reference frequency.
Accordingly, we define the `applied' detuning $x_0$ as
\begin{equation}\label{eqn:def_applied_detuning}
	x_0 = \frac{\nu - \frzero}{\frzero}.
\end{equation}
The concepts of \emph{applied} and \emph{realised} detuning will prove particularly useful in the next section.

\subsection{Point of zero realised detuning}\label{sec:zero_detuning}

The point of zero realised detuning, $x = y=0$, occurs when the measurement frequency is equal to the resonant frequency of the device despite the parametric changes present.
If a shunt resonator behaves purely linearly, the point of zero-detuning in the frequency-sweep can be identified from one of the following conditions:
(i) The transmission phase is (and crosses through) zero.
(ii) The transmission gain $T=|S_{21}|^2$ is minimised.
(iii) The reflection factor $R=|S_{11}|^2$ is maximised.
For linear resonators with S-parameters different from (\ref{eqn:mkid_s11}) and (\ref{eqn:mkid_s21}), equivalent conditions can be determined that will depend on $\Gamma_{mn}$ and $K_{mn}$ in (\ref{eqn:s_matrix_general}).

In the case of a non-linear resonator, we must look for the point in the sweep where $y(\nu, p) = 0$.
Here, the readout frequency is equal to the resonant frequency of the underlying resonance.
We will now show that aspects of the conditions (i)--(iii) carry over to distorted, and even hysteretic, cases.
Again we will assume the measurement arrangement of Section \ref{sec:origin}, and that the S-parameters of the device under test are given by (\ref{eqn:mkid_s11}) and (\ref{eqn:mkid_s21}).
The same methods can be applied to other types of device to derive equivalent conditions.

Consider the phase-shift on transmission through the non-linear resonator, as given by the argument of $S_{21}$.
The distorted curve is generated from (\ref{eqn:mkid_s21}) by varying $\fres$ and $\Qi$ with readout frequency, keeping both real.
(\ref{eqn:mkid_s21}) is such that $\text{Arg}[S_{21}] = 0$ if and only if $x = y = 0$; therefore, even in the case of a distorted curve, we know that the detuning is zero anywhere the transmission phase is zero, i.e. (i) still holds.

A possible source of confusion occurs experimentally when a device exhibits switching.
For example, the green dashed curve in Figure \ref{fig:resonance_circle_distortion} appears to pass through zero near $y_0 \approx -1.2$, but in actual fact the device is merely changing state, and the response is discontinuous: $y \neq 0$.
In practice, it should be easy to identify such cases because they coincide with similar discontinuities in $R$ and $T$.

To determine the stationary points of $R$ and $T$ for a non-linear resonator, we must calculate their derivatives with respect to the readout frequency.
It follows from (\ref{eqn:pdiss_full}) and (\ref{eqn:s11_and_s21_p_notation}) that
\begin{equation}\label{eqn:R_from_p}
	R = \frac{p}{2 \eta}
\end{equation}
and
\begin{equation}\label{eqn:T_from_p}
	T = 1 - p - R.
\end{equation}
Taking the total derivatives of (\ref{eqn:R_from_p}) and (\ref{eqn:T_from_p}) with respect to $\nu$ and then using the chain rule we obtain
\begin{equation}\label{eqn:deriv_R}
	\frac{dR}{d\nu} = \frac{1}{2\eta} \biggl[
		1 - \frac{p}{\eta} \frac{d\eta}{dp}
	\biggr] \frac{dp}{d\nu}
\end{equation}
and
\begin{equation}\label{eqn:deriv_T}
	\frac{dT}{d\nu} = -\frac{1}{2\eta} \biggl[
		1 + 2 \eta - \frac{p}{\eta} \frac{d\eta}{dp}
	\biggr] \frac{dp}{d\nu},
\end{equation}
where we have suppressed the dependence of $\eta$ on $p$ in the notation for convenience.

By taking the total derivative of (\ref{eqn:p_steady_state}) with respect to $\nu$, we can obtain the follow condition involving $dp/dv$
\begin{equation}\label{eqn:deriv_p_wrt_dv_eq1}
	\frac{dp}{d\nu} =
	\frac{p}{\eta} \biggl[ 1 - (1 + \eta) p \biggr] \frac{d\eta}{dp} \frac{dp}{dv} -
	\frac{4 p^2 y}{\eta} \frac{dy}{d\nu}.
\end{equation}
However, it also follows by partial differentiation that
\begin{equation}\label{eqn:y_total_derivative}
	\frac{dy}{d\nu} = \left( \frac{\partial y}{\partial \nu} \right)_p
	+ \left( \frac{\partial y}{\partial p} \right)_\nu \frac{dp}{d\nu}.
\end{equation}
Using (\ref{eqn:y_total_derivative}) to substitute for $dy/d\nu$ in (\ref{eqn:deriv_p_wrt_dv_eq1}) and then solving the resulting equation for $dp/dv$, we obtain
\begin{equation}\label{eqn:deriv_p_wrt_v_final}
	\frac{dp}{d\nu}
	= -\frac{4 p^2 \kappa y}{\eta} \left(\frac{\partial y}{\partial \nu} \right)_p
\end{equation}
where
\begin{equation}\label{eqn:def_kappa}
	\kappa^{-1} = 1
	- \frac{p}{\eta} \bigl[ 1 - (1 + \eta) p \bigr] \frac{d\eta}{dp}
	+ \frac{4 p^2 y}{\eta} \left(\frac{\partial y}{\partial p} \right)_\nu.
\end{equation}
According to (\ref{eqn:deriv_R}), (\ref{eqn:deriv_T}) and (\ref{eqn:deriv_p_wrt_v_final}), the derivatives can therefore be written as
\begin{equation}\label{eqn:deriv_R_final}
	\frac{dR}{d\nu}
%	= -\frac{2 \kappa p^2 y}{\eta^2}\left[
%	1 - \frac{p}{\eta} \frac{d\eta}{dp}
%	\right]
%	\left(\frac{\partial y}{\partial \nu} \right)_p
	= -8 \kappa y R^2 \left[
	1 - 2 R \frac{d\eta}{dp}
	\right]
	\left(\frac{\partial y}{\partial \nu} \right)_p
\end{equation}
and
\begin{equation}\label{eqn:deriv_T_final}
	\frac{dT}{d\nu}
%	= \frac{2 \kappa p^2 y}{\eta^2} \left[
%	1 + 2 \eta - \frac{p}{\eta} \frac{d\eta}{dp}
%	\right]
%	\left(\frac{\partial y}{\partial \nu} \right)_p
	= 8 \kappa y R^2 \left[
	1 - 2 (R + \eta) \frac{d\eta}{dp}
	\right]
	\left(\frac{\partial y}{\partial \nu} \right)_p.
\end{equation}
(\ref{eqn:deriv_R_final}) and (\ref{eqn:deriv_T_final}) indicate that $R$ and $T$ are stationary with respect to the sweep frequency at the point of zero-detuning of a non-linear resonator $y=0$, as for a linear device.

To evaluate the nature of the stationary point in each case we need to take a further derivative and evaluate the result at $y=0$.
Differentiating (\ref{eqn:deriv_R_final}) using the chain rule, discarding terms proportional to $y$ and noting that $dp/dv = 0$ at $y=0$, we obtain
\begin{equation}\label{eqn:second_deriv_R_wrt_freq}
	\left( \frac{d^2 \! R}{d\nu^2} \right)_{y=0}
	= -8 \kappa R^2 \left[ 1 - 2 R \frac{d\eta}{dp} \right]
	\left(\frac{\partial y}{\partial \nu} \right)_p^2
\end{equation}
\begin{equation}\label{eqn:second_deriv_T_wrt_freq}
	\left( \frac{d^2 \! T}{d\nu^2} \right)_{y=0}
	= 8 \kappa R^2 \left[ 1 - 2 (R + \eta) \frac{d\eta}{dp} \right]
	\left(\frac{\partial y}{\partial \nu} \right)_p^2.
\end{equation}
(\ref{eqn:second_deriv_R_wrt_freq}) and (\ref{eqn:second_deriv_T_wrt_freq}) indicate that $R$ is still minimised and $T$ is maximised at zero realised detuning provided the content of each square bracket is positive.
Violation of the latter conditions requires non-linear dissipation, because $d\eta / dp$ would need to be significantly different from zero.

(\ref{eqn:deriv_R_final}) and (\ref{eqn:deriv_T_final}) also show that $R$ and $T$ can also be stationary if the contents of the square bracket in each expression are zero.
Unlike for a linear resonator, we can therefore no longer automatically assume that \emph{any} stationary point in $R$ and $T$ is a point of zero realised detuning.
However, notice that the contents of the square bracket can only be zero for one or other of (\ref{eqn:deriv_R_final}) and (\ref{eqn:deriv_T_final}) at any time.
Therefore if $R$ and $T$ are stationary simultaneously, or the phase is also zero, we can still identify the point as corresponding to zero realised detuning.

Being able to identify the point of zero realised detuning using the conditions above is particularly convenient for parameter extraction, even under highly nonlinear conditions.
Most obviously, we know that if the point of zero realised detuning is at measurement frequency $\nu$, then
\begin{equation}\label{eqn:fres_at_zero_detuning}
	\fres(\Pdiss) = \nu.
\end{equation}
However, it follows from (\ref{eqn:p_res}) and (\ref{eqn:s11_and_s21_p_notation}) with $y=0$ that we can also calculate $\Qi (\Pdiss)$ and $\Pdiss$ from the S-parameters at the zero realised detuning point using
\begin{equation}\label{eqn:qi_at_zero_detuning}
	\frac{\Qi(\Pdiss)}{\Qc}
	= \frac{-S_{11}}{1 + S_{11}} = \frac{1 - S_{21}}{S_{21}}
\end{equation}
and
\begin{equation}
	\frac{\Pdiss}{\Pread}
	= -2(S_{11} + |S_{11}|^2)
	=  2( S_{21} - |S_{21}|^2).
\end{equation}
Thus the internal quality factor and dissipated power follow from measurements of the scattering parameters at the point of zero realised  detuning, which are real, even for a nonlinear device.
This technique can be used to great effect (Section \ref{sec:measurement_scheme}).

\subsection{Other stationary points}\label{sec:other_stationary_points}

It is instructive to consider the other cases where $R$ and $T$ can be stationary with frequency, as these might, potentially, be confused experimentally with the case $y=0$.
For both $R$ and $T$, the only other circumstance when this can occur is when the contents of the square brackets in (\ref{eqn:deriv_R}) and (\ref{eqn:deriv_T}) are zero.
In the case of $R$, this requires
\begin{equation}\label{eqn:r_stat_alt_condition}
	\frac{2p}{\eta} \frac{d\eta}{dp} = 1,
\end{equation}
which corresponds to the situation where the change in $p/\eta$ due to the change in readout frequency is cancelled out by the corresponding change in $\eta$ due to nonlinear behaviour.
It is straightforward to show that for a simple power model given by $\eta = \alpha p^n$, (\ref{eqn:r_stat_alt_condition}) can only be satisfied if $n=1$.
Furthermore, when $n=1$ the condition is actually satisfied for all $p$, and so $R$ becomes independent of readout frequency.
This behaviour would be easily distinguished from the case where $y=0$.

Similarly, in the case of $T$ we require
\begin{equation}\label{eqn:t_stat_alt_condition}
	1 + 2 \eta - \frac{p}{2 \eta} \frac{d\eta}{dp} = 0.
\end{equation}
For the power law model used above, this condition can be satisfied at a spot power $p = p_*$ where
\begin{equation}\label{eqn:t_stat_spot_power}
	p_* = \sqrt[n]{\frac{n-2}{4 \alpha}},
\end{equation}
provided $n > 2$.
However, the case where $n > 2$ is a very strong nonlinearity, which we will see in Section \ref{sec:nlm_exp_greater_than_one} produces a high level of distortion of the resonance shape.
As a result, it is unlikely we would confuse a stationary point resulting from this effect with one resulting from realising zero detuning.

\subsection{Kinetic inductance}\label{sec:accessing_zero_detuning}

To this point the analysis has been general, making no assumptions about the origins of the physical mechanisms that cause the resonant frequency and line width to depend on readout power, and perhaps other variables such as temperature.
In superconducting films, kinetic inductance introduces a reactive nonlinearity.
Kinetic inductance is the circuit-theoretic representation of energy stored in the inertial motion of Cooper pairs.
It has the beneficial effect that distributed resonators based on superconducting films are physically smaller than resonators based on normal metals.
However, for large currents $I$, the kinetic inductance is nonlinear:
\begin{equation}\label{eqn:kin_ind_nonlin}
    L = L_{0} \left[ 1 + \left( \frac{I}{I_{\ast 1}} \right)^{2}  + \left( \frac{I}{I_{\ast 2}} \right)^4 + \cdots \right],
\end{equation}
where $I_{\ast 1}$ and $I_{\ast 2}$ are scaling currents.
This nonlinearity can be used to create superconducting devices, such as travelling wave parametric amplifiers\,\cite{eom2012wideband}, but in the context of resonators, it leads to a redistribution of frequency points on the resonance circle, as shown by the green crosses in the left plot of Figure \ref{fig:resonance_circle_distortion}, and can cause hysteretic switching, as shown in the right plot.

\begin{figure}
\centering
\includegraphics{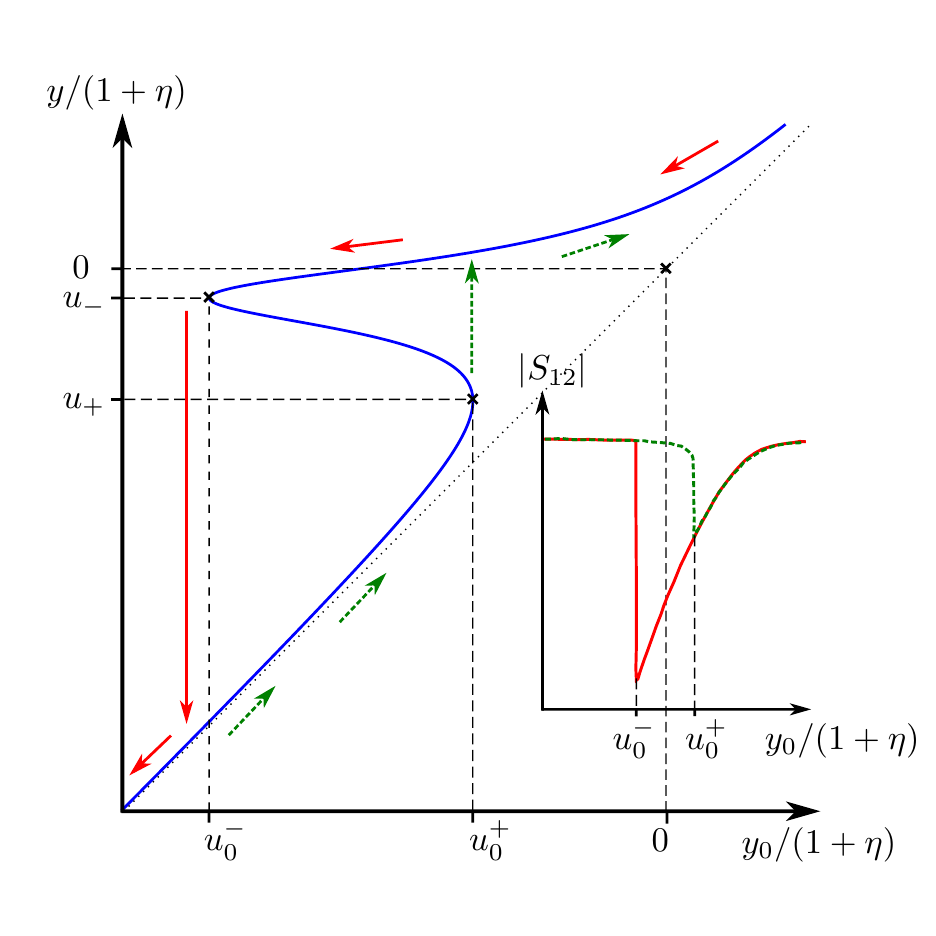}
\caption{\label{fig:reactive_nl_y_vs_y0}
Realised detuning $y$ as a function of zero-power detuning $y_0$; see also Figure 2 of Swenson\,\cite{swenson2013operation}.
Blue (solid) line shows the solution of (\ref{eqn:swenson_y_equation}) for $a = 5$ and the dotted line $y = y_0$ for comparison.
Red arrows indicate the trajectory of the resonator in the $(y_0, y)$-plane when $y_0$ is swept in the negative direction from a large, positive, starting value.
Dashed-green arrows show the opposing case where $y_0$ is instead swept in the positive direction from a large negative value.
Lines with matching format in the inset show the variation in $|S_{21}|$ with $y_0$ in each case ($\Qt = \Qc$).
}
\end{figure}

Strictly, the inclusion of nonlinear inductance leads to complicated periodic forms for the voltage, current and inductance, but using the expression $\fres = (L C)^{-1/2}$; keeping only the quadratic term in (\ref{eqn:kin_ind_nonlin}); concentrating on those spectral components that are at the same frequency as the readout tone; and using the stored energy as a proxy for the square of the average current; we find that
\begin{equation}\label{eqn:swen_mod}
	\fres(U) = \frzero \left[ 1 - U / \Uckin \right],
\end{equation}
where $\frzero$ is the resonance frequency in the low-energy limit, and $\Uckin$ scales the size of the nonlinear effect.
Swenson's model\,\cite{swenson2013operation} assumes that the resonant frequency decreases linearly with stored energy $\Ures$, and has been found experimentally to provide a good description of certain non-linear reactive behaviour in superconducting resonators\,\cite{swenson2013operation,mccarrick2014horn,semenov2016coherent}.
The internal quality $\Qi$ is, according to the model, constant, and so the system has only reactive nonlinearity.
Substituting (\ref{eqn:swen_mod}) into (\ref{eqn:def_detuning_x}) and then Taylor expanding assuming small $U/\Uckin$, we find the detuning $y$, as defined by (\ref{eqn:def_y}), becomes
\begin{equation}
	y = \frac{\Qc [ \fread - \fres(U)]}{\fres(U)}
	\approx y_0 + \frac{\Qc U}{\Uckin}
	= y_0 + \frac{\Qc \Qi \Pread}{2 \pi \frzero \Uckin} p,
\end{equation}
where $y_0 = \Qc x_0$ is the applied detuning relative to $\frzero$, as defined by (\ref{eqn:def_applied_detuning}).

Not only does kinetic inductance redistribute the frequency points on the resonance circle, it can cause hysteretic switching.
One consequence is that a point of zero detuning may not be found anywhere during a swept frequency measurement.
We can illustrate this effect as follows.

Using (\ref{eqn:p_res}) to substitute for $p$, we obtain
\begin{equation}\label{eqn:swenson_y_equation}
	y = y_0
	+ \frac{(1 + \eta)^3 a}{(1 + \eta)^2 + 4y^2},
\end{equation}
where $a = \Qt^3 \Pread / \pi \frzero \Qc \Uckin$ is Swenson's non-linearity parameter in our notation.
Note that $y$ as defined in Swenson's paper\,\cite{swenson2013operation} corresponds to $y/(1 + \eta)$ in our formulation, as they measure linewidths relative to $\Qt$ rather than $\Qc$.
For given applied detuning $y_0$, (\ref{eqn:swenson_y_equation}) can be solved to find the realised detuning $y$ and entered into (\ref{eqn:mkid_s11}) and (\ref{eqn:mkid_s21}).

For values of $a > 4 \sqrt{3} /9$ (Appendix \ref{sec:swenson_additional}), $y$ is not monotonic in $y_0$ and different resonant curves are obtained depending on whether the measurement frequency is swept up or down.
This is illustrated for $a = 5$ in Figure \ref{fig:reactive_nl_y_vs_y0}, which reproduces part of Figure 2 from \cite{swenson2013operation}.
When the readout frequency is swept up, the resonator follows the trajectory in the $(y_0, y)$-plane indicated by the dotted-green arrows.
Critically, at $y_0 = (1 + \eta) u^+_0$ the value of $y$ jumps discontinuously from $(1 + \eta) u_+$ to some higher value.
Similarly, when the readout frequency is swept down it follows the trajectory shown by the solid-red arrows and $y$ jumps discontinously from $u_-$ to some smaller value at $y_0 = y_0^-$.
The inset of Figure \ref{fig:reactive_nl_y_vs_y0} shows the corresponding curves of $|S_{21}|$ versus $y_0/(1+\eta)$.

It is possible for $y$ to skip through the point of zero-detuning in one of these jumps; whether it does so depends on the values of $u_+$ and $u_-$, as well as the value of $u$ afterwards.
It can be seen from Figure \ref{fig:reactive_nl_y_vs_y0} that the jump points correspond to stationary points of $y_0$ as a function of $y$.
Taking the derivative of (\ref{eqn:swenson_y_equation}) with respect to $y$ and then setting $dy/dy_0$ equal to zero, we find $u_+$ and $u_-$ must satisfy
\begin{equation}\label{eqn:up_and_un_equation}
	-\frac{1}{8a} = \frac{u_\pm}{(1 + 4 u_\pm^2)^2}.
\end{equation}
Since the non-linearity parameter is always positive, (\ref{eqn:up_and_un_equation}) implies that $u_+$ and $u_-$ are both always negative.
Viewing Figure \ref{fig:reactive_nl_y_vs_y0} from the perspective of $y$ as a function of $y_{0}$, it is apparent that $y$ is always guaranteed to pass through zero detuning ($y=0$) on a downward sweep from well above resonance: extrema then appear in the magnitudes of the scattering parameters.
However, on an upward sweep, the resonator may jump to a positive or negative value of detuning, depending on the precise shape, and extrema will only appear in the former case.
Note that if instead the resonant frequency increases with stored energy (e.g. as observed in the higher temperature data in \cite{de2014evidence}), this behaviour would be reversed.
The shape of measured hysteretic resonance curves therefore change in specific ways, revealing key information about the underlying nonlinearities.
Effects of this kind are seen routinely in experimental resonance curves.
Some further useful results concerning the locations of the switching points are described in Appendix \ref{sec:swenson_additional}.

\section{Two-level systems}\label{sec:tls}

In low-temperature superconducting resonators, two mechanisms are found to be dominant sources of dissipative nonlinearity.
The first relates to the presence of Two Level Systems (TLS) in deposited bulk and unintended surface oxides (such as SiO$_{2}$), and the second relates to heating and pair breaking in the films that make up the transmission lines.

TLSs occur in amorphous insulating materials where local configurational changes in the atoms that make up the material lead to changes in stored mechanical energy.
According to the low-temperature TLS model, a system can tunnel between one configurational state and another, introducing a new low-energy degree of freedom into the dynamical behaviour\,\cite{anderson1972anomalous,phillips1972tunneling,phillips1987two}.
The TLS model has been highly effective at describing the low temperature behaviour of heat capacity, sound speed, and acoustic attenuation.
If, additionally, the TLSs have an electric dipole moment, they can contribute significantly to the electromagnetic properties, leading to an enhanced dielectric constant, which may have a dissipative part due to energy being carried away by elastic waves.
TLSs have proven extremely successful at explaining empirical data for detuning, loss and noise in thin-film superconducting resonators \,\cite{gao2008semiempirical,pappas2011two,gao2008physics,sage2011study,burnett2016analysis}.

In most practical devices, the exact nature and locations of the TLSs are not known, and it is usual to imagine some density of TLSs having an assumed energy distribution.
Detailed theoretical models exist for the real and imaginary parts of the dielectric constant, but for our purposes the important features are as follows:
(i) The dielectric constant has two parts, one of which is due to the coupling of the TLSs to the phonon system, which acts as a thermalising reservoir, and the other is caused by resonant transitions between TLS states.
(ii) The first relaxation process gives a complex dielectric constant that is independent of field strength, and leads to damped linear-resonator behaviour.
(iii) The second resonant process has a real part that depends only weakly on field strength, giving a weak reactive nonlinearity, and a lossy imaginary part that depends strongly on field strength, giving a strong dissipative nonlinearity.
For a sufficiently strong field, the resonant energy states can be driven to have equal populations, and the losses become zero.
For parameterisation, it is sufficient to know that
\begin{equation}\label{eqn:tls_qtls_from_u}
	\Qtls^{-1} = \frac{\Qtlsmin^{-1}}{\sqrt{1 + U / \Uctls}},
\end{equation}
where $U$ is the energy stored in the electric field, $\Uctls$ characterises the energy at which the TLSs saturate, and $\Qtlsmin^{-1}$ characterises the maximum power loss.
This expression should be compared with the functional form in (\ref{eqn:swen_mod}), where $\Uckin$ characterises the energy at which nonlinear inductance starts to become significant.
In resonators of practical importance $\Uckin > \Uctls$, allowing for some intermediate range of readout power where linear resonator behaviour can be found.
This is usually regarded as the `sweet spot',  $\Uckin > U >  \Uctls$, for device operation.

In (\ref{eqn:tls_qtls_from_u})  $U$ can be replaced by either the internal resonator power $P_\text{int} = 2 \pi \fres U$ (different to $\Pread$ or $\Pdiss$)\,\cite{gao2008physics} or the square $|E|^2$ of some measure $E$ of the electric field strength in the capacitive part of the resonator\,\cite{mazin2010thin}.
All three forms of (\ref{eqn:tls_qtls_from_u}) are equivalent, but we choose to work with $U$ because it can be defined in a geometry independent manner, with all details of the design of the device absorbed into $\Uctls$.

Consider a resonator where TLSs are the only source of nonlinearity.
Using the definition of the internal quality factor,
\begin{equation}\label{eqn:def_qi}
	U = \Qi \frac{\Pdiss}{2 \pi \fres},
\end{equation}
and (\ref{eqn:p_res}), it can be shown that the total energy $U$ stored in the resonator is
\begin{equation}\label{eqn:tls_resonator_energy}
	U = \frac{2 \Qt^2}{\Qc} \frac{1}{1 + (2 \Qt x)^2} \frac{\Pread}{2 \pi \fres},
\end{equation}
where $\Pread$ is the incident readout power.
This expression is true for both the series and shunt single-pole resonant circuits.
It is immediately obvious from (\ref{eqn:tls_qtls_from_u}) and (\ref{eqn:tls_resonator_energy}) that $\Qtls$ depends on $\Pread$.
However, the functional form of the relationship is difficult to obtain.
(\ref{eqn:tls_resonator_energy}) cannot simply be used to calculate $U$ as an input to (\ref{eqn:tls_qtls_from_u}), as $\Qt$ is itself a function of $\Qtls$.
The two equations must instead be solved as a pair of nonlinear simultaneous equations.

Experimental studies to verify (\ref{eqn:tls_qtls_from_u}) have avoided this difficulty by exploiting the fact that the value of $\Qt$ measured to calculate $\Qtls$ can be used to convert $\Pread$ to $U$ (or actually usually $P_\text{int}$)\,\cite{gao2008physics,mazin2010thin}.
However, there are many situations where is it valuable to calculate $\Qt$ as a function of $\Pread$, for example when explaining experimental data directly or when designing a device.
To our knowledge this problem has not been addressed in the literature, so we will do so in the next section.

\subsection{Large signal model and numerical solution}\label{sec:tls_model}

Assume that the nonlinear behaviour of TLSs only affects the dissipative response of the resonator, so the detuning $x$ is fixed.
Using (\ref{eqn:tls_qtls_from_u}), the total resonator quality factor is
\begin{equation}\label{eqn:tls_tls_inv_qr}
	\Qt^{-1} = \Qc^{-1} + \Qimax^{-1} + \Qtlsmin^{-1} / \sqrt{1 + u},
\end{equation}
where $u = U / \Uctls$ and $\Qimax^{-1}$ represents any other sources of internal loss that ultimately limit the achievable quality factor.
It is convenient to rewrite (\ref{eqn:tls_tls_inv_qr}) in the form
\begin{equation}\label{eqn:tls_qr_from_alpha}
	\Qt = \frac{\Qtmin}{1 - r \alpha},
\end{equation}
where $\Qtmin = (\Qc^{-1} + \Qimax^{-1} + \Qtlsmin^{-1})^{-1}$ and $\Qtmax = (\Qc^{-1} + \Qimax^{-1})^{-1}$ are, respectively, the smallest and largest values $\Qt$ can take, $r = \Qtmin  / \Qtlsmin = (\Qtmax - \Qtmin) / \Qtmax$, and
\begin{equation}\label{eqn:tls_master_eqn}
	\alpha = 1 - \frac{1}{\sqrt{1 + u}}
\end{equation}
measures the state of the TLS system under applied power. $0 \leq r, \alpha \leq 1$ by definition. $\alpha = 0$ and $1$ correspond to the limits where the TLSs are fully unsaturated and saturated, respectively.
To determine the steady-state behaviour, we must solve for $\alpha$ at the readout power level given known $x$, $r$, $\fres$, $\Qc$ and $\Qtmin$.

Substituting (\ref{eqn:tls_resonator_energy}) into (\ref{eqn:tls_master_eqn}), we find that the determination of $\alpha$ can be posed as the fixed-point problem
\begin{equation}\label{eqn:tls_fixed_point_problem}
	\alpha = f (\alpha)
\end{equation}
for
\begin{equation}\label{eqn:tls_def_f}
	f(\alpha) = 1 - \frac{(1 - r \alpha)}{\sqrt{(1 - r \alpha)^2 + \chi(\alpha) \ytls}},
\end{equation}
where 	
\begin{equation}\label{eqn:tls_def_ytls}
	\ytls = \frac{\Pread}{\Pctls},
\end{equation}
is a dimensionless nonlinearity parameter,
\begin{equation}\label{eqn:tls_def_ptls}
	\Pctls = \frac{\pi \fres \Qc \Uctls}{\Qtmin^2}
\end{equation}
is a scale power and
\begin{equation}\label{eqn:tls_def_chi}
	\chi (\alpha) = \frac{(1 - r \alpha)^2}{(1 - r \alpha)^2 + (2 \Qtmin x)^2}
\end{equation}
is the quantity normally referred to as the detuning efficiency\,\cite{zmuidzinas2012superconducting}.

By definition, $\ytls \geq 0$ and $0 \leq \chi \leq 1$.
The advantage of putting the problem in this form is that certain fixed-point theorems can be applied to its solution.
A full discussion is given in Appendix \ref{sec:tls_proofs}, but the key results can be summarised as follows.
First, we can show that (\ref{eqn:tls_fixed_point_problem}) always has one unique solution satisfying the physical constraint $0 < \alpha < 1$, which precludes the existence of hysteretic effects due to the action of TLS alone.
Second, we can show that the iterative sequence defined by
\begin{equation}\label{eqn:tls_iterative_sequence}
	\alpha_{n+1} = f(\alpha_n)
\end{equation}
always converges to this solution in the limit $n \rightarrow \infty$ provided the sequence is started from $a_0 = 0_+$.

\subsection{Simulated behaviour}\label{sec:simulated_tls_behaviour}

\begin{figure}
\centering
\includegraphics[width=8cm]{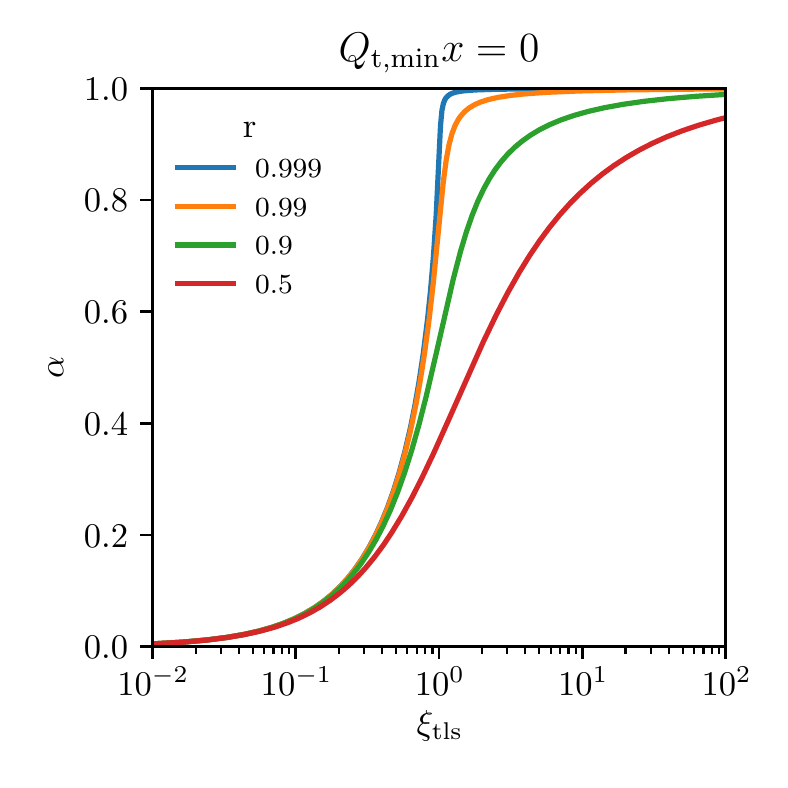}
\caption{\label{fig:tls_alpha_vs_y} Plot of $\alpha$, which characterises the state of the TLS system in our formulation, as a function of $\ytls$ for different values of the ratio $r = \Qtmin / \Qtlsmin$.
Zero detuning is assumed.}
\end{figure}

\begin{figure}
\centering
\includegraphics[width=8cm]{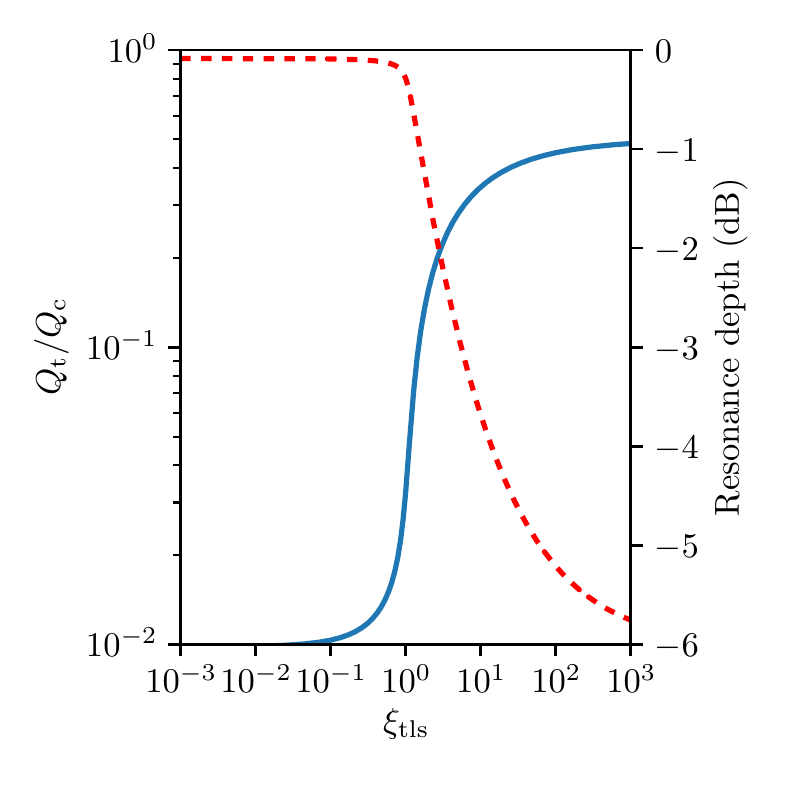}
\caption{\label{fig:tls_res_depth_vs_pr} $\Qt / \Qc$ (blue solid line, left-hand $y$-axis) and transmission at zero-detuning in dB (red dashed line, right-hand $y$-axis) as function of $\ytls$.
$\Qc$ = $\Qimax$ = $10^5$ and $\Qtlsmin$ = $10^3$ ($r = 0.98$).}
\end{figure}

Figure \ref{fig:tls_alpha_vs_y} shows calculated curves of $\alpha$ versus $\ytls$ for a range of values of $r$ at zero detuning ($x=0$).
It can be seen that $\alpha$ becomes an increasingly sharp step-like function as $r \rightarrow 1$, which corresponds to the physical limit where TLS loss dominates internal loss at low power.
The step change in $\alpha$ occurs at $\ytls=1$, or equivalently, using (\ref{eqn:tls_def_ytls}), when the readout power level is near the critical power level $\Pctls$.

The implications for device behaviour can be seen in Figure \ref{fig:tls_res_depth_vs_pr}, which shows calculated values of $\Qt / \Qc$ (blue solid line, left $y$-axis) and resonance depth (red dashed line, right $y$-axis) as function of $\ytls$ for $x=0$.
The assumed values of the various $Q$-factors are given in the figure caption and $r \approx 0.98$.
The sharp increase in $\alpha$ at $\ytls \approx 1$ leads to a rapid increase in $\Qt / \Qc$ when the readout power is raised above some threshold value.
Equivalently, this can be seen as a very rapid increase in the depth of the resonance from nearly 0\,dB to -6\,dB over an order of magnitude change in $\ytls$ (or, equivalently, applied readout power).
Under certain experimental conditions this behaviour gives rise to a `switch-on' effect: the resonator is obscured by the noise floor of the system and appears absent until the readout power is increased above a threshold, at which point the depth increases rapidly and the resonance curve `turns on'.
We have seen this striking behaviour in a number of our own devices having high levels of TLS loss.

\begin{figure}
\centering
\includegraphics[width=8cm]{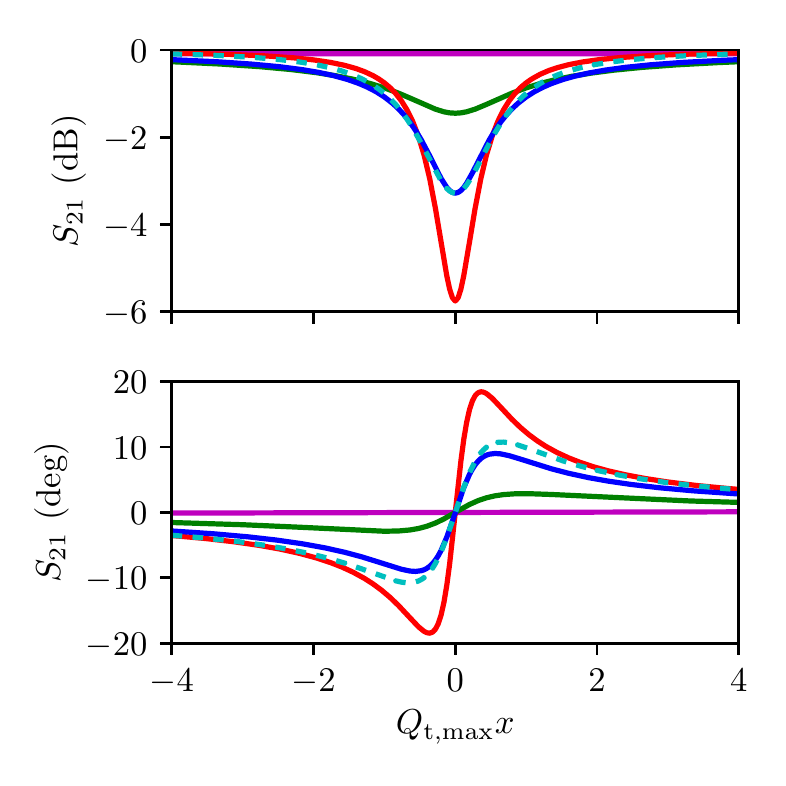}
\caption{\label{fig:tls_res_shape_vs_pr} Resonance curves as a function of $\ytls$ for the same resonator parameters as Figure \ref{fig:tls_res_depth_vs_pr}.
The solid lines show data from the full model, showing the resonance deepening as $\ytls$ increases.
The values of $\ytls$ for the different lines are as follows: $\ytls = 0.1$ (magenta); $\ytls =2$ (green); $\ytls = 7$ (blue); and $\ytls = 1000$ (red). 
The dashed cyan line shows the response of a linear device, with $\Qt / \Qc$ chosen to match that of the non-linear device at zero detuning in the case $\ytls = 7$.
}
\end{figure}

This switch-on behaviour and associated distortion is illustrated in Figure \ref{fig:tls_res_shape_vs_pr}, which shows calculated resonance curves for different values of $\ytls$ assuming the same device parameters as in Figure \ref{fig:tls_res_depth_vs_pr}.
The cases $\ytls = 0.1$ (magenta line) and $10^3$ (red line) correspond to the limits where the TLS are fully unsaturated and fully saturated, respectively (as can be seen from Figure \ref{fig:tls_res_depth_vs_pr}).
In the case $\ytls = 0.1$, the resonance curve is too shallow to be seen on the graph scales we have used.
In these regimes the behaviour of the amplitude and phase as a function of frequency is indistinguishable from that of a linear device, as we will see explicitly when we consider the resonance curves in the Argand plane.

For $\ytls = 2$ (green line), the TLS are just starting to saturate and the resonance curve becomes visible.
The solid blue lines show the components of $S_{21}$ for what is effectively the mid-point in the saturation process: $\ytls = 7 $.
The dashed cyan line shows the ideal linear response calculated using (\ref{eqn:mkid_s21}) and a value of $\Qt$ calculated from the depth of the fully modelled response for the green line at zero detuning.
As can be seen, the dashed curves fall off more slowly then the full model, which is consistent with a reduction in Q in the full model as the energy stored in the resonator falls and the saturation state of the TLSs decreases.
Even in this worst case regime, the distortion in amplitude is relatively slight, although there is a stronger effect in the phase.
Such distortion may still affect the fitting of (\ref{eqn:mkid_s21}) to experimental curves; in particular, we might expect a good fit to either the width or depth, but not both simultaneously.

\begin{figure}
\centering
\includegraphics[width=8cm]{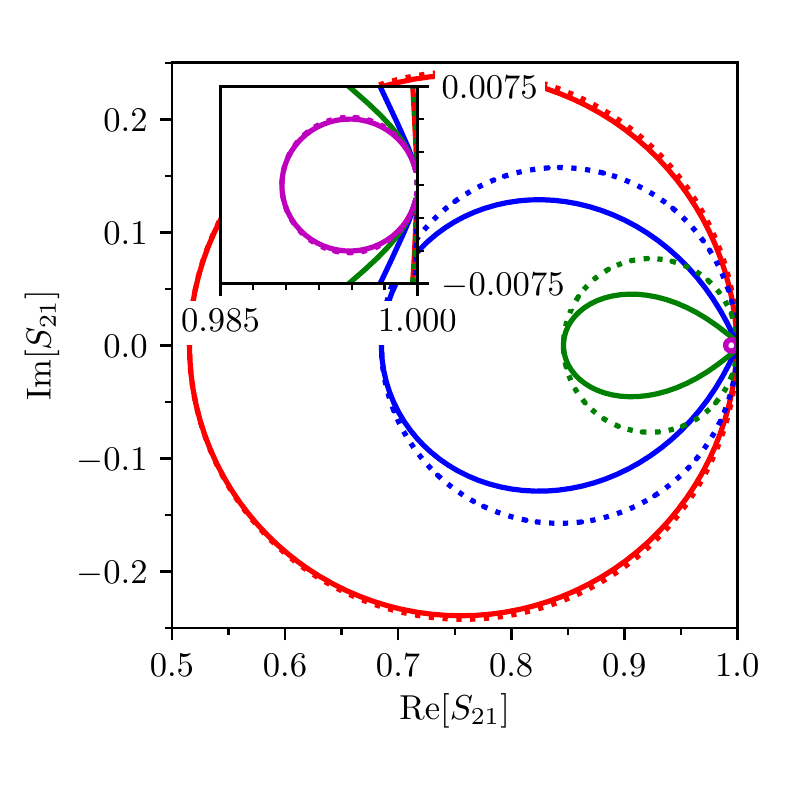}
\caption{\label{fig:tls_iq_circles}
Resonance `circles' as a function of $\ytls$ for the same resonator parameters as Figure \ref{fig:tls_res_shape_vs_pr}.
Solid lines show modelled response, while the dashed lines show the corresponding circles assuming the $Q$-factors measured on resonance.
The values of $\ytls$ for the different lines are as follows: $\ytls = 1000$ (red); $\ytls = 7$ (blue); $\ytls =2$ (green); and $\ytls = 0.1$ (magenta)
}
\end{figure}

\begin{figure}
\centering
\includegraphics[width=8cm]{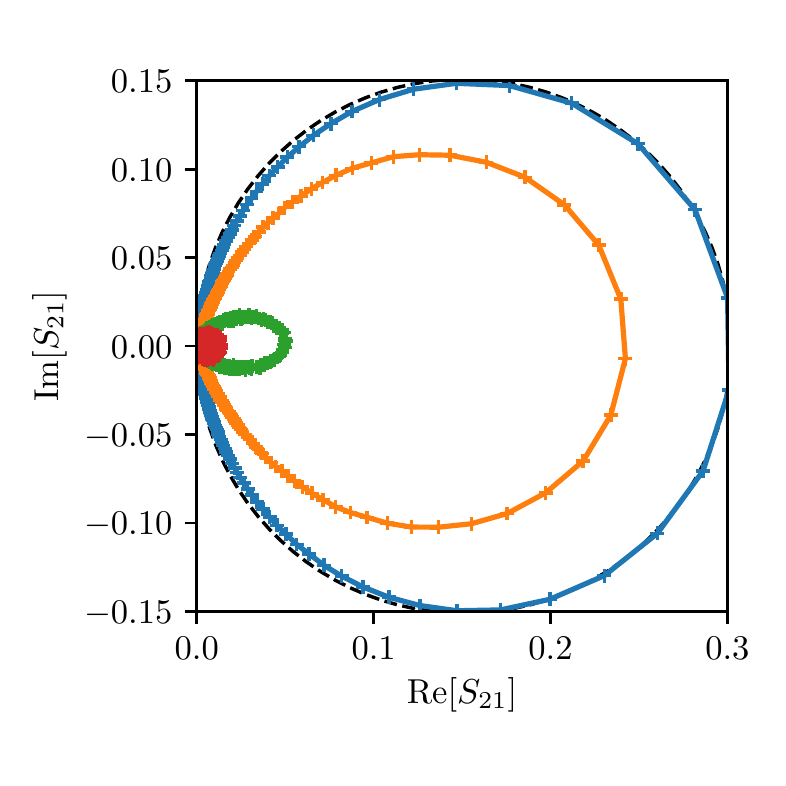}
\caption{\label{fig:tls_iq_circles_exp}
Experimental data demonstrating the behaviour illustrated in Figure \ref{fig:tls_iq_circles}.
The different colours indicate different power levels; in each case the crosses indicate measured data points and lines of matching colour have been drawn between them for emphasis.
The applied readout power increases in 10\,dBm increments going from red to green, green to orange and orange to blue.
}
\end{figure}

The distortion of the resonance curve is most apparent in the Argand plane.
Figure \ref{fig:tls_iq_circles} shows calculated response in the Argand plane, using the same parameters and colours of Figure \ref{fig:tls_res_shape_vs_pr}.
The resonance curves form circles when the TLSs are either fully unsaturated or saturated throughout the frequency sweep, but in the intermediate range ($\ytls = 2$ and $\ytls = 7$) a `teardrop' shape is seen.
The formation of this teardrop shape is a result of $\Qtls$, and therefore the radius of the resonance circle (Section \ref{sec:nl_in_argand_plane}), decreasing rapidly as the measurement signal is tuned off resonance and the energy stored in the resonator decreases.
We have seen this effect in many of our microstrip devices and Figure \ref{fig:tls_iq_circles_exp} shows typical measured data.
In this case the device was a half-wave resonator and the microstrip with a 2\,$\mu$m wide, 400\,nm thick, Nb trace, 500\,nm sputtered SiO$_2$ dielectric layer and a 150\,nm thick Nb ground plane ($\Tc$ $\approx$ 8\,K).
The measurements were taken at 110\,mK.

\section{Quasiparticle heating}\label{sec:quasiparticle_heating}

In superconducting resonators, Ohmic dissipated readout power can have a marked effect on resonance curves, even when the readout frequency is well below the pair-breaking energy gap of the material.
Multiple sequential photon absorption events, starting with a thermal population, can pump the quasiparticle system into a highly non-equilibrium state, which loses energy to the phonon system of the underlying material.
The application of readout power effects both the energy distribution of the quasiparticles \emph{and} their number density.
The complex processes by which the quasiparticle and phonon energy spectra are modified in the presence of sub-gap photons have been studied at the microscopic level by Goldie\,\cite{goldie2012non}, and the predictions have subsequently been found to be in excellent agreement with experimental results\,\cite{de2014evidence}.
In the context of resonator dynamics, a key observation is that the consequential macroscopic behaviour can be described by a reduced model where the quasiparticles are ascribed an effective temperature above their physical temperature.
The power dissipated by the readout signal effectively heats the quasiparticles\,\cite{de2010readout}, and an equilibrium state is formed when the heating power is balanced by the cooling power flow to the phonons\,\cite{goldie2012non,guruswamy2015nonequilibrium}.
This {\em electrothermal model} has been used to account for both large-signal\,\cite{de2010readout,thompson2013dynamical} and small-signal\,\cite{thomas2015electrothermal} device behaviour.

Here we introduce an alternate, but equivalent, macroscopic model based on the Rothwarf-Taylor equations\,\cite{rothwarf1967measurement}, which replaces the effective temperature with the total quasiparticle number.
We will show this model is closely related to the electrothermal model, but is advantageous for our application because it allows approximate forms for $\Qi$ as a function of $\Pread$ to be obtained easily for comparison with experimental results.

\subsection{Description of the model}\label{sec:rt_model_description}

Our primary aim is to calculate how the internal quality factor $\Qqp$ varies with applied readout power.
In the limit where the operating temperature is well below the critical temperature of the superconductor $\Tc$ (usually taken as $T / \Tc < 0.1$), and the resonant frequency is well below the pair-breaking frequency, Mattis-Bardeen theory predicts $\Qqp$ to be inversely proportional to the number density of quasiparticles $\nqp$ in the active part of the resonator (see Appendix \ref{sec:qqp_as_a_function_of_n} for proof, also noted by McCarrick\,\cite{mccarrick2014horn}).
For our purposes, it will be convenient to express this relationship in the form
\begin{equation}\label{eqn:qp_qi_from_nqp}
	\Qqp = \frac{\nstar \Qc}{\nqp}.
\end{equation}
The scaling factor $\nstar$ absorbs the effects of temperature, frequency and resonator geometry, and can be recognised as the quasiparticle density at which $\Qqp = \Qc$.
Choosing $\Qc$ as the characteristic scale for $\Qqp$ will be advantageous later when we consider how the actual power dissipated in the resonator relates to the applied readout power $\Pread$.

The `active part' of the resonator in this context is determined by the current distribution.
By definition, (\ref{eqn:def_qn}), $\Qqp$ is inversely proportional to the total Ohmic power dissipation in certain volumes, $V_1, V_2, \dots V_N$, of the superconducting device.
In the temperature-frequency range of interest, the resistivity of a superconductor is small and approximately proportional to the local quasiparticle density (Appendix \ref{sec:qqp_as_a_function_of_n}).
Hence we expect
\begin{equation}\label{eqn:q_spatial_integral}
    \Qqp \propto \sum_{i=1}^N \int_{V_i} \nqp |\mathbf{J}|^2 d\tau
\end{equation}
where $\mathbf{J}$ is the local induced current density and $\int_{V_i} \dots d\tau$ denotes the volume integral over $V_i$.
(\ref{eqn:q_spatial_integral}) indicates $\Qqp$ will be predominantly determined by $\nqp$ in the region of highest current density; for example, nearest the shorted end of a quarter-wave resonator.
Similarly, most of the power will be dissipated in the same region.
Consequently, it is sufficient to only consider the evolution of $\nqp$ in high-current regions when determining $\Qqp$ to first order.

To determine how $\nqp$ depends on $\Pread$, our starting point is the Rothwarf-Taylor equations\,\cite{rothwarf1967measurement} in the form
\begin{gather}
	\frac{d\nqp}{dt} = \frac{2}{\tpb} \nomega - \Rrc \nqp^2
	\label{eqn:qp_rt_qp} \\
	\frac{d\nomega}{dt} = -\frac{1}{\tpb} \nomega + \frac{\Rrc}{2} \nqp^2
		- \frac{1}{\tl} \left[ \nomega - \nomegath \right] + \Gamma_\text{r}.
	\label{eqn:qp_rt_ph}
\end{gather}
$\nqp$ is the number density of quasiparticles in the active volume of the resonator, $\nomega$ is the number density of pair-breaking phonons in the same volume, and $\nomegath$ is the value of $\nomega$ in thermal equilibrium, $\Gamma_\text{r} = 0$ (no forcing).
$\tpb$ is the pair-breaking time, $\Rrc$ is the quasiparticle recombination rate and $\tl$ is the lifetime of a pair-breaking phonon in the absence of interactions with the quasiparticle system.
$\Gamma_\text{r}$ is the rate at which pair-breaking phonons are generated by the readout signal.

We are interested in the steady-state behaviour, and so we set $\partial_t \nqp$ and $\partial_t \nomega$ equal to zero.
(\ref{eqn:qp_rt_ph}) can then be used to eliminate $\nomega$ in (\ref{eqn:qp_rt_qp}), and we find the steady-state value of $\nqp$ must satisfy
\begin{equation}\label{eqn:qp_eqlbrm_1}
	\Rrc \nqp^2
	= \frac{2 \tau_\text{l}}{\tpb}
	\left[ \Gamma_\text{r} + \frac{1}{\tl} \nomegath \right].
\end{equation}
A further simplification  is possible if recognise that $\nqp$ must be equal to the expected value thermal $\nqpth$ when $\Gamma_\text{r} =0$, so $2 \nomegath / \tl = \Rrc \nqpth^2$.
(\ref{eqn:qp_eqlbrm_1}) can therefore be re-expressed as
\begin{equation}\label{eqn:qp_eqlbrm_2}
	\Rrc \left[ \nqp^2 - \nqpth^2 \right]
	= \frac{2 \tl}{\tpb} \Gamma_\text{r},
\end{equation}
where it has been shown that
\begin{equation}\label{eqn:qp_thermal_qp_density}
	\nqpth = 2 N_0 \sqrt{2 \pi k_\text{b} T \Delta} e^{-\Delta / k_\text{b} T},
\end{equation}
where $T$, $\Delta$ and $N_0$ are respectively the temperature, gap energy and single spin density of states at the Fermi surface of the superconductor\,\cite{thomas2015electrothermal}.

As of yet we have not said anything about how the generation rate is related to quasiparticle number density and readout power.
As a first approximation, we assume
\begin{equation}\label{eqn:qp_gamma_r}
	\Gamma_\text{r} \approx \frac{\eff P_\text{qp}}{V},
\end{equation}
where $P_\text{qp}$ is the total power dissipated in the quasiparticle system, $V$ is the volume of the active part of the resonator and $\eff$ is a generation efficiency.

To relate $P_\text{qp}$ to the applied readout power $\Pread$, we must consider both the effects of the resonator circuit and the division of power between the different loss mechanisms.

Let
\begin{equation}\label{eqn:qp_qi_in_heating_model}
	\Qi^{-1} = \Qqp^{-1} + \Qo^{-1},
\end{equation}
where $\Qo$ collects together all other internal losses in the resonator.
By definition,
\begin{equation}\label{eqn:qp_pqp}
	P_\text{qp} = \frac{\Qi}{\Qqp} \Pdiss
\end{equation}
where $\Pdiss$ is the total power dissipated in the resonator given by
\begin{equation}\label{eqn:qp_power_dissipated}
	\Pdiss = \frac{2\Qi \Qc}{(\Qi + \Qc)^2}
	\frac{1}{1 + (2 \Qt x)^2} \Pread.
\end{equation}
Due to the way in which experimental data is often taken, we assume $x = 0$ in the subsequent analysis.
However, it is also straightforward to apply the model for finite $x$ and to also account for distortion caused by the resonant frequency changing with $\nqp$ ($\delta x \propto \nqp^{-1}$), but we shall not do so here.

Combining (\ref{eqn:qp_gamma_r})--(\ref{eqn:qp_power_dissipated}),
\begin{equation}\label{eqn:qp_pqp_full}
	\Gamma_\text{r} = \frac{\Qqp^{-1} \Qc^{-1}}
	{(\Qo^{-1} + \Qqp^{-1} + \Qc^{-1})^2}
	\frac{2 \eff \Pread}{V}.
\end{equation}
(\ref{eqn:qp_qi_from_nqp}) can be used to rewrite (\ref{eqn:qp_pqp_full}) in terms of quasiparticle number densities instead of quality factors.
Doing so, and substituting the result into (\ref{eqn:qp_eqlbrm_2}), yields
\begin{equation}\label{eqn:qp_master_equation}
	\Rrc \left[ \nqp^2 - \nqpth^2 \right]
	= \frac{4 \eff \tl \Pread}{\tpb V}
	\frac{\nstar \nqp}{(\nstar \left[ 1 + \Qc / \Qo \right] + \nqp)^2},
\end{equation}
which must be solved to find $\nqp$ in equilibrium.
(\ref{eqn:qp_master_equation}) can be rearranged into a quartic equation in $\nqp$, and must generally be solved numerically, as will be discussed in subsequent sections.
However, first consider the relationship between this model and previous models of quasiparticle heating in superconducting resonators.

\subsection{Relation to effective temperature models}\label{sec:effective_temperature}

(\ref{eqn:qp_eqlbrm_1}) suggests that the dynamics of $\nqp$ near equilibrium can be described by the rate equation
\begin{equation}\label{eqn:qp_rate_equation}
	\frac{d \nqp}{dt}
	\approx \frac{2 \tl}{\tpb} \Gamma_\text{r}
	- \Rrc \left[ \nqp^2 - \nqpth^2 \right],
\end{equation}
with the implication being that recombination dominates the loss mechanisms.
For operating temperature $T_0$ well below the critical temperature of the superconductor and small enough $\nqp$, the total energy $U_\text{qp}$ of the quasiparticle system is $\approx \nqp V \Delta$: see Thomas\,\cite{thomas2015electrothermal}.
Further, let us use (\ref{eqn:qp_thermal_qp_density}) to assign an effective temperature $T_\text{qp}$ to the quasiparticles which makes the expected thermal value equal to the nonequilibrium value $\nqp$.
Multiplying (\ref{eqn:qp_rate_equation}) through by $V \Delta$ and using (\ref{eqn:qp_thermal_qp_density}) to replace $\nqp$ and $\nqpth$ with expressions in terms of effective temperatures results in the energy balance equation
\begin{equation}\label{eqn:qp_rate_eqn_for_u}
	\frac{dU_\text{qp}}{dt} = P_\text{in}
	- \kappa_0^2 \left[ T_\text{qp} e^{-2 \Delta / k_\text{b} T_\text{qp}}
	- T_0 e^{-2 \Delta / k_\text{b} T_0} \right]
\end{equation}
for $P_\text{in} = 2 \tl \Gamma_\text{r} \Delta / \tpb$ and $\kappa_0 = 2 \Rrc n_0 V \sqrt{2 \pi k_\text{b} \Delta^2}$.
(\ref{eqn:qp_rate_eqn_for_u}) reproduces the effective temperature and superconducting cooling curve model developed in the series of papers \cite{de2010readout, goldie2012non, thompson2013dynamical, guruswamy2015nonequilibrium, thomas2015electrothermal}.
The model introduced in this paper can therefore be viewed as a reformulation of the existing microscopic electrothermal model, but the approach taken here is favoured because it simplifies some of the subsequent mathematics.

It is interesting also to compare the model proposed here with that from Section 5.6.4 of Zmuidzinas\,\cite{zmuidzinas2012superconducting}.
His model is based on the empirical observation that the quasiparticle relaxation time $\tau$ saturates at $\tmax$ as $T/\Tc$ is reduced.
Given an assumed dependence
\begin{equation}\label{eqn:qp_limit_lifetime}
	\tau = \frac{\tmax}{1 + \nqp / \ntau},
\end{equation}
Zmuidzinas derives, in our notation, a total generation rate
\begin{equation}\label{eqn:qp_zmuidzinas_rate}
	\Gamma = \frac{2 \tl}{\tpb} \Gamma_\text{r}
	- \Rrc \left[ \nqp^2 - \nqpth^2 \right] - \Rrc \ntau \left[ \nqp - \nqpth \right],
\end{equation}
where $\ntau \tmax = 1 / \Rrc$.
This differs from the total generation rate in (\ref{eqn:qp_rate_equation}) by the term linear in $\nqp$, so we expect the models to diverge in the regime $\nqp \approx \nqpth$.
Since we will be mainly concerned with the regime where $\nqp \gg \nqpth$, we will not dwell on this difference.
However, in Section \ref{sec:overcoupled_device} will show that in our model $\nqpth$ limits at $\nqpmin$ as the temperature is reduced, as a result of readout power heating.
This gives rise to the behaviour described by (\ref{eqn:qp_limit_lifetime}), without the need to impose a limited relaxation time.

\subsection{Full solution}\label{sec:full_solution}

\begin{figure}
\centering
\includegraphics{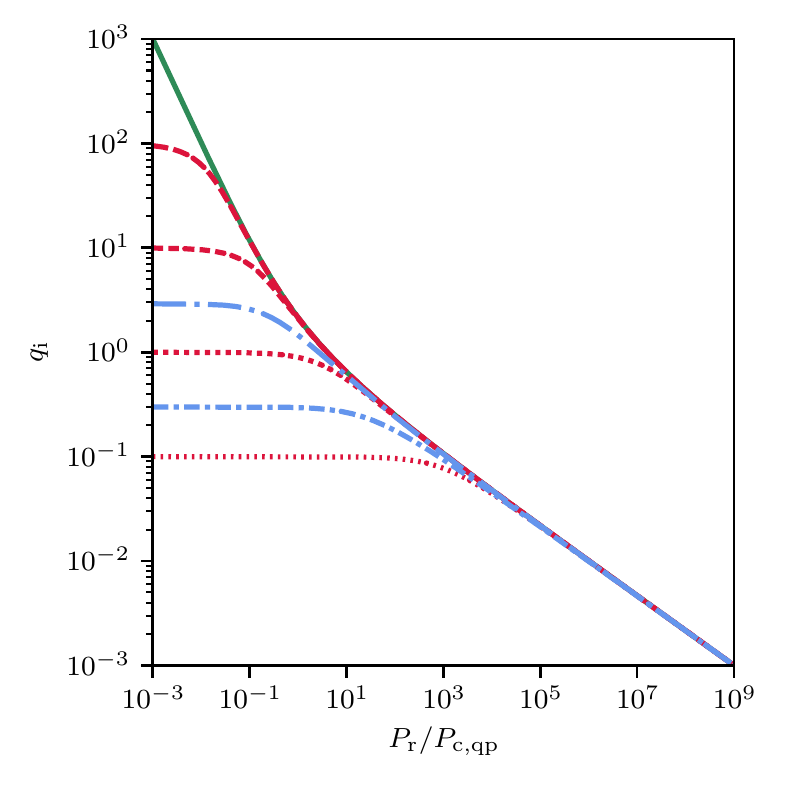}
\caption{\label{fig:numerical_qi_vs_p}
Plot of $\qi$ as a function of $\Pread / \Pcqp$, as found by solving (\ref{eqn:qp_quartic_equation_for_n}).
The dashed (red) lines show the effect of reducing the thermal quasiparticle density when other losses are fixed and small ($\qo = 10^6$).
As the dash length increases, $\qqpth$ increases through the sequence of values 0.1, 1, 10 and 100.
The dot-dash (blue) lines show the case where $\qqpth = 100$ and $\qo$ increases from 0.1 (short dashes) to 1 (long dashes).
The solid (green) shows the case where $\qqpth = \qo = 10^7$, i.e. losses from both mechanisms are small at low powers.
}
\end{figure}

(\ref{eqn:qp_master_equation}) can be rewritten as the quartic equation
\begin{equation}\label{eqn:qp_quartic_equation_for_n}
\begin{aligned}
	0 =& n^4 + 2 \left(1 + \qo^{-1} \right) n^3
	+ \left[ \left( 1 + \qo^{-1} \right)^2 - \qqpth^{-2} \right] n^2 \\
	&- \left[ 2 \qqpth^{-2} \left( 1 + \qo^{-1} \right) + \Pread / \Pcqp \right] n
	- \qqpth^{-2} \left( 1 + \qo^{-1} \right)^2
\end{aligned}
\end{equation}
for normalised variables $n = \nqp / \nstar$ and $\qqpth = \Qqpth / \Qc$, where $\Qqpth = \nstar \Qc / \nqpth$ is the quality factor expected from thermal quasiparticles alone and
\begin{equation}\label{eqn:def_pcqp}
    \Pcqp = \tpb \nstar^2 \Rrc V / 4 \eff \tl
\end{equation}
is a scaling power.
(\ref{eqn:qp_quartic_equation_for_n}) can be solved numerically using a root-finding algorithm and selecting for the roots that satisfy the physical requirements that $n$ must be real and greater than or equal to zero.
In all the simulations described here, this procedure yielded a single solution.

Figure \ref{fig:numerical_qi_vs_p} shows calculated values of $\qi$ as a function of $\Pread / \Pcqp$ for a range of values of $\Qqpth$ and $\Qo$.
For readout powers well above $\Pcqp$, all the curves lie on top of each other.
In this regime the behaviour is dominated by the population of quasiparticles excited by the readout power, and so differences in other losses or thermal quasiparticle number have no influence.
As the power is reduced, each continues along a common path until $\Qi$ saturates at the smaller of either $\Qqpth$ or $\Qo$, with no apparent difference in the shape of the curve depending on the source of the limiting value.
In the sections that follow we will derive simplified forms for $\Qqp$ as a function of applied readout power in a number of relevant cases.

\subsection{Behaviour of an over-coupled device}\label{sec:overcoupled_device}

$\Qc \ll \Qi$ for an overcoupled device.
This requires $\nqp \ll \nstar$ if quasiparticle losses dominate the internal losses in the resonator, which follows from (\ref{eqn:qp_qi_from_nqp}).
(\ref{eqn:qp_master_equation}) can then be approximated by
\begin{equation}\label{eqn:qp_governing_equation_overcoupled}
	\Rrc \left[ \nqp^2 - \nqpth^2 \right] =
	\frac{2 \eff \tl \Pread}{\tpb V} \frac{\nqp}{\nstar},
\end{equation}
which can be rearranged into a quadratic equation in $\nqp$ and solved analytically.
Only the solution
\begin{gather}
	\nqp = \frac{\nqpmin}{2} + \sqrt{\left( \frac{\nqpmin}{2} \right)^2 + \nqpth^2}
	\label{eqn:qp_overcoupled_nqp} \\
	\nqpmin = \frac{4 \eta \tl \Pread}{\tpb \nstar V \Rrc}
	\label{eqn:qp_nqpmin}
\end{gather}
satisfies the physical requirement that $\nqp \geq 0$.

(\ref{eqn:qp_overcoupled_nqp}) indicates that $\nqp$ will not decrease indefinitely as device temperature is reduced in the presence of a readout signal.
Instead it reaches a minimum value $\nqpmin$ corresponding to an excess population maintained by readout power dissipated in the device.
At first this may seem counter-intuitive; if the losses are decreased to zero, where does the dissipated power to both initiate and then maintain this population come from?
The answer is the influence of the resonator circuit.
The right-hand side of (\ref{eqn:qp_governing_equation_overcoupled}) indicates the electrical behaviour of the resonator provides positive feedback in the overcoupled-limit: a small increase in $\nqp$ produces an increase in generation rate, tending to further increase $\nqp$.
Consequently, the generation of a few quasiparticles --for example, by a noise process or optical  event-- is sufficient to start the process.
The loss from this process then provides sufficient dissipated power to sustain the population.

\subsection{Behaviour of an under-coupled device}\label{sec:under_coupled_device}

A device is over-coupled if $\Qi \ll \Qc$.
If quasiparticles again dominate the internal loss in the resonator, (\ref{eqn:qp_qi_from_nqp}) now requires $\nqp \gg \nstar$ and (\ref{eqn:qp_master_equation}) can be approximated by
\begin{equation}\label{eqn:qp_governing_equation_undercoupled}
	\Rrc \left[ \nqp^2 - \nqpth^2 \right] =
	\frac{2 \eff \tl \Pread}{\tpb V} \frac{\nstar}{\nqp}.
\end{equation}
The quasiparticle term on the right-hand side is inverted compared with the over-coupled case, (\ref{eqn:qp_governing_equation_overcoupled}), and so the resonator power provides negative feedback: an increase in $\nqp$ reduces the rate at which quasiparticles are generated.

(\ref{eqn:qp_governing_equation_undercoupled}) can be rearranged into a cubic equation and an analytic solution is possible, however we will make a further simplification.
We will assume $\nqp \gg \nqpth$, which would correspond experimentally to the case where the measured $\Qi$ is much smaller than would be predicted on the basis of an assumed thermal population of quasiparticles.
We therefore approximate $\nqp^2 - \nqpth^2 \approx \nqp^2$, in which case (\ref{eqn:qp_governing_equation_undercoupled}) yields
\begin{equation}\label{eqn:qp_nqp_high_power_limit}
	\nqp =\left( \frac{2 \eff \tl \nstar \Pread}{\tpb \Rrc V} \right)^{1/3}.
\end{equation}
Substituting (\ref{eqn:qp_nqp_high_power_limit}) and $\Pread = G P_\text{VNA}$ into (\ref{eqn:qp_qi_from_nqp}) and taking the logarithm of the result, we obtain the prediction
\begin{equation}\label{eqn:qp_log_qi_undercoupled}
	\log_{10} \Qi =
	\frac{1}{30} \log_{10} \left(
		\frac{2 \eta \tl \nstar G}{\tpb \Rrc V}
		\times 1\,\left[ \text{mW} \right]
	\right) - \frac{1}{30} P_\text{VNA} \left[ \text{dBm} \right],
\end{equation}
which may be readily compared with experimental data.

\subsection{Comparison with experiment}\label{sec:qp_comparison_with_exp}

\begin{figure}
\centering
\includegraphics[width=8cm]{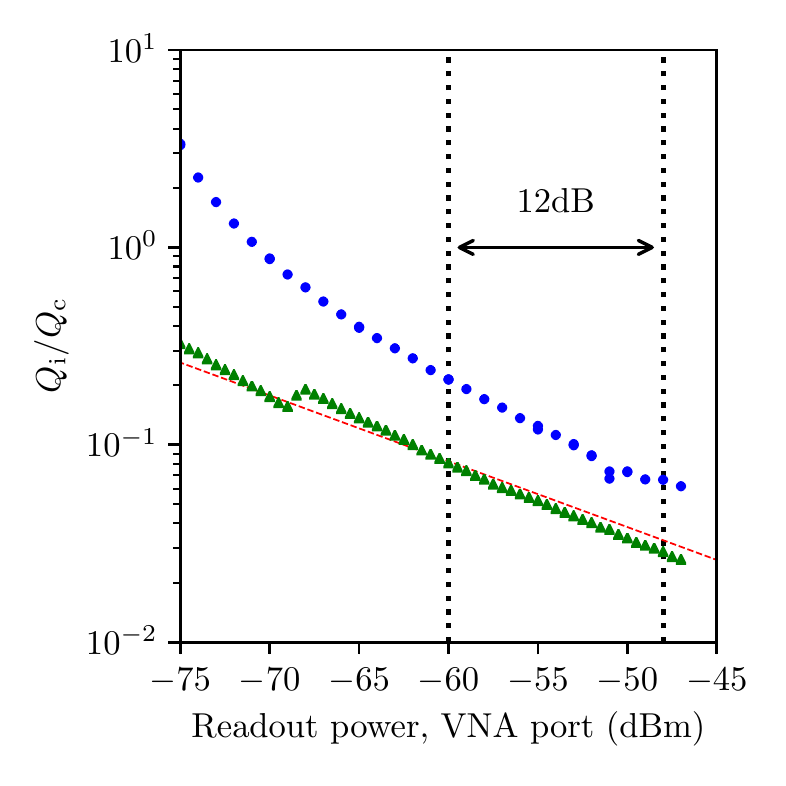}
\caption{\label{fig:example_data_qi_vs_pvna}
Measured values of normalised internal quality factor $\Qi / \Qc = \qi$ as a function of applied readout power for NbN-SiO$_2$ microstrip resonators in the regime where quasiparticle heating is expected.
Device geometry is described in Section \ref{sec:qp_comparison_with_exp}.
The blue circles and green triangles show data for devices with measured coupling quality factors of $3.6 \times 10^4$ and $1.4 \times 10^5$ respectively.
The red dashed line shows a fit of (\ref{eqn:qp_log_qi_undercoupled}) to the green triangles with the intercept as a free variable.
}
\end{figure}

Figure \ref{fig:example_data_qi_vs_pvna} shows measurements of the normalised internal quality factor as a function of readout power of two superconducting resonators in a regime where quasiparticle heating is expected.
In both devices the resonator is a quarter-wave length of superconducting microstrip.
One end of this line is shorted, and the other is lightly capacitively to a readout line, yielding a circuit similar to the top panel of Figure \ref{fig:different_resonator_geometries}.
The microstrip comprises a 2.5\,$\mu$m wide and 200\,nm thick, reactively sputtered, NbN ground plane, 550\,nm thick RF sputtered SiO$_2$ dielectric layer and a 400\,nm thick NbN ground plane.
Measurements on a monitor sample from the depositions indicate the resistivity of the NbN is approximately 300\,$\mu\Omega$cm, and the superconducting critical temperature 10.8\,K.
The devices were from two chips designed to differ in coupling strength; $\Qc$ was measured as $3.6 \times 10^4$ for the device represented by the blue circles and $1.4 \times 10^5$ for that represented by the green triangles.
The measurements were taken at 100\,mK using the method described in Section \ref{sec:measurement_scheme}.
VNA power is a proportional measure of the readout power $\Pread$ applied at the device.

The device with the lower value of $\Qc$ (green triangles) is under-coupled at even the lowest readout powers and should, therefore, be in the regime discussed in Section \ref{sec:under_coupled_device}.
The dashed red line in Figure \ref{fig:example_data_qi_vs_pvna} shows a fit of the straight-line model (\ref{eqn:qp_log_qi_undercoupled}) to the corresponding data with the intercept as a free variable.
The model can indeed be seen to provide a good account of the behaviour of $\qi$ with readout power.
As an additional test, we also attempted fitting both the gradient and intercept simultaneously using linear regression.
This gave a value for the reciprocal of the gradient of 26$\pm$0.3\,dBm, which is close to but slightly below the value 30\,dBm in (\ref{eqn:qp_log_qi_undercoupled}).
However, this is consistent with the fact the gradient of $1/30$\,dBm$^{-1}$ is the limit for very high powers and that  the actual gradient approaches it from above, as shown in Figure \ref{fig:numerical_qi_vs_p}.

The device with the higher value of $\Qc$ (blue circles) starts in the critically coupled regime, intermediate between the results of Sections \ref{sec:overcoupled_device} and \ref{sec:under_coupled_device}.
Consequently, there is no simplified expression to fit to the data.
However, the data is qualitatively similar to the prediction of Figure \ref{fig:numerical_qi_vs_p}, with the gradient of the line seen to decrease as applied power increases.
Further, the value of the gradient is approximately correct in the two limits:
i) At the point of critical coupling $\Qc = \Qi$, it is about $1/20$\,dBm$^{-1}$, in agreement with the Figure \ref{fig:numerical_qi_vs_p}.
ii) At higher powers and under-coupling the blue circles begin to trace a line nearly parallel the dashed red line, indicating the behaviour is tending to (\ref{eqn:qp_log_qi_undercoupled}).

The discussion above already indicates the quasiparticle heating model gives a reasonable account of the behaviour of each device individually.
However, we can go further and relate the values measured between devices.
The behaviour shown in Figure \ref{fig:numerical_qi_vs_p} is universal, therefore any horizontal offset between the two sets of points in Figure \ref{fig:example_data_qi_vs_pvna} should result purely from the difference in the scaling powers $\Pcqp$ of the devices.
Given they are of similar design and composition, (\ref{eqn:def_pcqp}) indicates the ratio of the scaling powers should be proportional to the ratio of $n_*^2$ for the two devices.
However, by definition $n_*$ is inversely proportional to $\Qc$ under the same conditions.
Consequently, given the measured values of $\Qc$ we should expect $\Pctls$ for the strongly coupled device (blue circles) to be approximately fifteen times that for the more weakly coupled device (green triangles).
In turn, this translates into a predicted shift of 12\,dB between the two datasets in Figure \ref{fig:example_data_qi_vs_pvna} at similar values of $\Qi / \Qc$.
As can be seen, this is a very good description of what is actually observed.

\section{Power law models}\label{sec:power_law_models}

\subsection{Model and method of solution}\label{sec:power_law_model_desc}

In previous sections, we considered the effects of TLS and quasiparticle heating.
For these specific mechanisms we are able to calculate the functional form of the quality factor with respect to dissipated power, and explain what was seen experimentally.
Often, however, we are in the converse situation: we have measured a set of distorted resonance curves and we would like to determine, or at least infer, the functional form of the underlying physical process.
In this section, we will describe a power-law model, which helps to develop an intuition for functional forms that produce specific shapes in distorted resonance circles.

Let $\Qi$ be decomposed into a power-independent contribution $\Qo$ and a contribution $\Qnl$ from nonlinear dissipative processes,
where
\begin{equation}\label{eqn:plm_qi}
	\Qi^{-1} = \Qo^{-1} + \Qnl^{-1}.
\end{equation}
Now assume that $\Qnl$ has a simple power-law form, dependent on the power $\Pnl$ dissipated by the nonlinear process:
\begin{equation}\label{eqn:plm_qnl}
	\Qnl = \Qc \left( \frac{\Pcpl}{\Pnl} \right)^{n/d},
\end{equation}
where $n$ and $d$ are positive integers (meaning that the exponent is always a rational number) and $\Pcpl$ is a parameter that determines the readout power level at which any nonlinear behaviour is seen.
In physical terms, this model describes a process where the dissipation increases with dissipated power; heating, for example.

We have assumed that $\Qnl$ depends directly on $\Pnl$ rather than the total power dissipated in the device, $P_\text{d}$, because this condition is expected to be more reflective of real processes.
For example, consider the quasiparticle heating model of Section \ref{sec:quasiparticle_heating}.
In this case, sub-gap readout photons are able to indirectly break pairs because the power $P_\text{qp}$ dissipated in the quasiparticle system is reprocessed into pair-breaking phonons.
Breaking pairs increases the quasiparticle number, which in turn increases dissipation and decreases $\Qqp$; hence $\Qqp$ decreases with $\Pnl$.
However, we would not expect power dissipated in the dielectric or elsewhere to have the same effect (at least in the absence of significant heating).
Thus the correct dependence is $\Qqp (P_\text{qp})$ in this case, not $\Qqp (P_\text{d})$.
A counter example, would be if a device is poorly thermally anchored to its refrigerator, and then all of the dissipated power would lead to a change in temperature, and loss.

The value of $\Pnl$ at a given value of $y$ and $P_\text{r}$ can be found as follows.
From (\ref{eqn:def_qn}) it follows that $\Pnl$ is related to the total power dissipated in the resonator by $\Pnl = \Qi P_\text{d} / \Qnl$, and so using (\ref{eqn:p_res}) and the notation of previous sections,
\begin{equation}\label{eqn:plm_pnl}
	\Pnl = \frac{2 \qnl^{-1}}
	{(1 + \qi^{-1})^2 + (2 y)^2} P_\text{r}.
\end{equation}
In the steady state, $\Pnl$ must satisfy (\ref{eqn:plm_pnl}) for $\Qi$ given by (\ref{eqn:plm_qi}) and (\ref{eqn:plm_qnl}).
This condition can be expressed as the fixed point problem
\begin{equation}\label{eqn:plm_fixed_point}
	\rho = h(\rho),
\end{equation}
where $\rho = \Pnl / \Pcpl$, $\rho_\text{r} = P_\text{r} / \Pcpl$ and
\begin{equation}\label{eqn:plm_h}
	h(x) = \frac{2 \rho_\text{r} x^{n/d}}
	{(1 + \qo^{-1} + x^{n/d})^2 + (2y)^2},
\end{equation}
which provides a way of calculating a set of possible values of $\Pnl$.

Although we could solve (\ref{eqn:plm_fixed_point}) by iteration, as in Section \ref{sec:tls_model}, here there is a better alternative.
The condition $x = h(x)$ can be rearranged into the form
\begin{equation}\label{eqn:plm_polynomial}
	\kappa^{(2n + d)}
	+ [ (1 + \qo^{-1})^2 + (2 y)^2 ] \kappa^{d}
	+ 2 (1 + \qo^{-1}) \kappa^{n + d}
	- 2 \rho_\text{r} \kappa^n
	= 0
\end{equation}
where $\kappa = x^{1 / d}$.	
It can now be seen that the fixed points of $h(x)$ correspond to the nth powers of the roots of the polynomial in $\kappa$ on the left-hand side of (\ref{eqn:plm_polynomial}).
As a result the full set of fixed points can be quickly found using a polynomial root-finding algorithm, which are common in mathematical software packages.
It also follows that $h(x)$ has at most $2n + d$ unique fixed points.

Given the set of fixed points, how can we determine which corresponds to the realised value of $\rho$?
As a first step, fixed points that correspond to unphysical solutions can be eliminated: as a normalised power, $\rho$ must be purely real and greater than or equal to zero.
If multiple possibilities remain, which fixed point is realised at the operating point will depend on the stability of the corresponding state and the history of the device.
Unstable states will not be realised in practice.
If multiple stable states remain, then how the device has been prepared becomes important.
For example, when a parameter is being swept, each time it changes the resonator will tend to move to which ever of the new states is closest to its previous state with respect to $\Pnl$.

Normally the stability of a state would be assessed in relation to some potential equation in the underlying physical model.
This is not possible here, and so we adopt as our stability condition the requirement that the iterative sequence $x_{n+1} = h(x_n)$ started near enough the fixed point $x = x_0$ will converge to $x_0$ as $n$ tends to infinity.
The physical motivation is that the iterative process mirrors how the resonator will move to the new operating point when a parameter is changed, or, perhaps more importantly, how it will move back to the state if perturbed from it.
The only difference is that, in reality, the process is continuous and limited by the dynamical times of the resonator.

The stability condition is equivalent to requiring $|h'(x)| < 1$ for $x_0 - \delta_- < x < x_0 + \delta_+$ for some $\delta_-$ and $\delta_+ > 0$, where $x_0$ is the fixed point and $h'(x) = dh/dx$.
As a result, it is impossible for the fixed point to correspond to a stable solution if $|h'(x_0)| \geq 1 $.
Differentiating (\ref{eqn:plm_h}), it is straightforward to show
\begin{equation}\label{eqn:plm_deriv_h}
	h'(x_0) =
	\begin{cases}
		\Bigl[ 1 - \frac{(1 + \qo^{-1} + x_0^{n/d}) x_0}
		{2 \rho_\text{r}} \Bigr] \frac{n}{d} & x_0 \neq 0 \\
		0 & x_0 = 0 \text{ and } n > d \\
		\frac{2 \rho_\text{r}}
		{(1 + \qo^{-1})^2 + (2y)^2} & x_0 = 0 \text{ and } n = d \\
		\rightarrow O(\infty) & x_0 \rightarrow 0 \text{ and } n < d.
	\end{cases}
\end{equation}
We see that $\rho = 0$ is never a stable state for finite $\rho_\text{r}$ if $n < d$.
As far as we can tell, $h(x)$ is a relatively well behaved function for $n \geq d$, so we make the assumption it is sufficiently smooth that if  $|h'(x_0)| < 1$ we can also find a small region around $x_0$ for which  $|h'(x)|$ is also $< 1$.
Hence the stability conditions become:
i) If $x_0 \neq 0$, stability requires
\begin{equation}\label{eqn:plm_non_zero_x0_stability_req}
	\left| 1 - \frac{(1 + \qo^{-1} + x_0^{n/d}) x_0}
		{2 \rho_\text{r}} \right| \frac{n}{d} < 1.
\end{equation}
ii) If $x_0 = 0$ and  $n > d$ then $x_0$ always corresponds to a stable state.
iii) If $x_0 = 0$ and $n = d$, then for stability requires
\begin{equation}\label{eqn:plm_order_1_stability}
	2 \rho_\text{r} < (1 + \qo^{-1})^2 + (2y)^2.
\end{equation}
iv) If $x_0 = 0$ and $n < d$, the corresponding state is always unstable.

Finally, it is useful to consider the limiting behaviour of the model when $y=0$ and $\rho_\text{r} \rightarrow \infty$.
This is relevant to measurements of resonance depth as a function of applied readout power.
In this limit we expect $x^{n/d}$ near the solution to be sufficiently large compared with other terms that we can make the approximation
\begin{equation}\label{eqn:plm_h_approx}
	h(x) \approx \frac{2 \rho_\text{r}}
	{x^{n/d}},
\end{equation}
in which case
\begin{equation}\label{eqn:plm_rho_large_pr}
	\rho \approx (2 \rho_\text{r})^{d / (n + d)}.
\end{equation}
The resulting expression for the depth of the resonance is
\begin{equation}\label{eqn:plm_depth_power_law}
	|1 - S_{21}(y = 0)| \approx (2 \rho_\text{r})^{-d / (n + d)},
\end{equation}
which is a simple power law.

\subsection{Power law exponent less than one}\label{sec:nlm_exp_less_than_one}

\begin{figure}
\centering
\begin{subfigure}{0.48\textwidth}
\centering
\includegraphics[width=\textwidth]{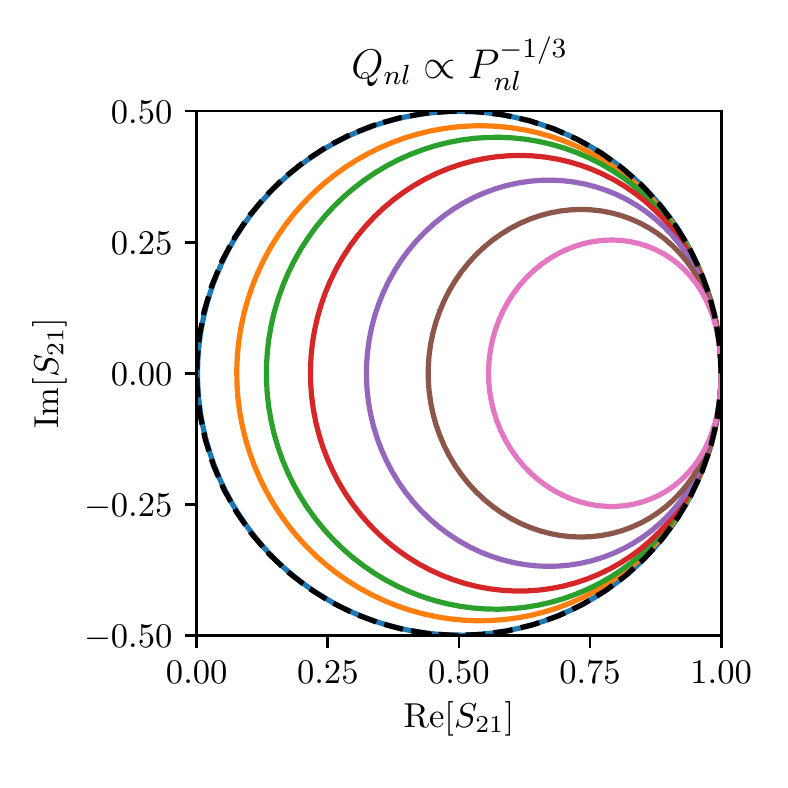}
\caption{\label{fig:argand_circle_plots_n_1_d_3}
Behaviour for $n=1$, $d=3$ and $\qo = 10^8$.
$\rho_\text{r} = 1 / 4^4 = 0.0039$ for the orange line and quadruples between lines as the circles get smaller, terminating in $\rho_\text{r} = 4$ for the pink line.
\vspace{1.2cm}
}
\end{subfigure}
\quad
\begin{subfigure}{0.48\textwidth}
\centering
\includegraphics[width=\textwidth]{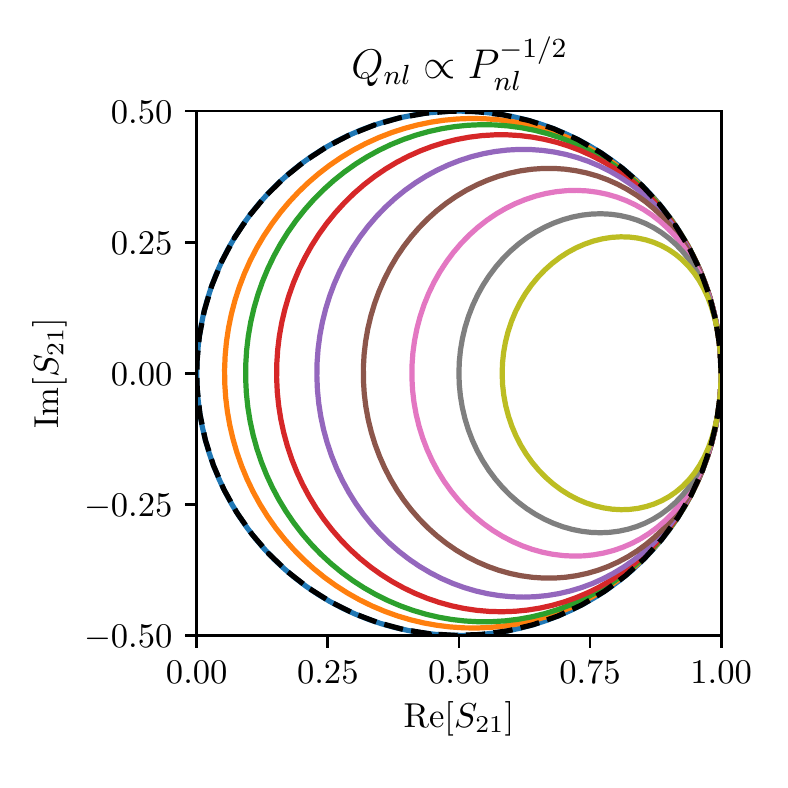}
\caption{\label{fig:argand_circle_plots_n_1_d_2}
Behaviour for $n=1$, $d=2$ and $\qo = 10^8$.
$\rho_\text{r} = 1 / 2^4 = 0.0625$ for the orange line and quadruples between lines as the circles get smaller, terminating in $\rho_\text{r} = 8$ for the yellow line.
Note that the range of values of $\rho_\text{r}$ shown is larger than in (a), i.e. the circle shrinks less rapidly as the applied power is increased.
}
\end{subfigure}
\caption{\label{fig:plm_argand_plots_root_exp}
Distorted resonance circles calculated using the model of Section \ref{sec:power_law_model_desc}.
In both plots the dashed black line shows a circle of radius $0.5$ centered on $S_{21} = 0.5$, which would be the expected behaviour of a highly over-coupled device ($\qi >> 1$).
}
\end{figure}

Figures \ref{fig:argand_circle_plots_n_1_d_3} and \ref{fig:argand_circle_plots_n_1_d_2} show simulated resonance `circles' in the Argand plane resulting from frequency sweeps at different readout power levels, for $n/d=1/3$ and $n/d=1/2$ with $\qo = 10^8$.
These illustrate typical behaviour when $n/d < 1$.
In both cases, the size of the resonance circle is observed to decrease with applied readout power.
At high powers the trajectory becomes distinctly non-circular and it is evident that it would not be possible to fit a single-pole model with fixed $\qi$ to the data.
Decreasing $d$ is observed to have two effects.
First, we see that the rate at which the size of the circle shrinks increases; in
Figure \ref{fig:argand_circle_plots_n_1_d_3} the difference in $\rho_\text{r}$ between neighbouring lines is a factor of four, while in Figure \ref{fig:argand_circle_plots_n_1_d_2} it is only a factor of two.
This is consistent with (\ref{eqn:plm_depth_power_law}).
Second, the circle is seen to become more asymmetric.
Finally, we draw attention to the fact that at high powers the radius of the circle is reduced at even high values of $y$.
This is a result of the fact the solution $\rho = 0$ is always unstable for $n/d 1$.
As we will see shortly, the behaviour is very different when $n/d > 1$.
	
\subsection{Power law exponent equal to one}\label{sec:nlm_exp_equals_one}	

In the case  $n= d$, yielding $\Qnl \propto \Pnl$, the model has an analytic solution.
(\ref{eqn:plm_polynomial}) reduces to the cubic equation
\begin{equation}\label{eqn:plm_cubic}
	\left\{ \rho^2
	+ 2 (1 + \qo^{-1}) \rho
	+ [(1 + \qo^{-1})^2 + (2y)^2 - 2 \rho_\text{r}]
	\right\} \rho = 0,
\end{equation}
with up to three unique solutions.
As factored it can be immediately seen that one solution is $\rho = 0$.
The other two solutions, $\rho = \rho_+$ and $\rho =\rho_-$, follow by solving the quadratic equation that results when the contents of the parentheses is set equal to zero, yielding
\begin{equation}\label{eqn:plm_rho_pm}
	\rho_\pm = -(1 + \qo^{-1}) \pm \sqrt{2 \rho_\text{r} - (2y)^2}.
\end{equation}

Of the three solutions, only $\rho = 0$ and $\rho = \rho_+$ correspond to possible physical states as $\rho_-$ is negative for all $\rho_\text{r}$ and $y$.
Further, $\rho_+$ is only positive if $\rho_\text{r}$ is greater than a threshold power $\rho_\text{t}$, where
\begin{equation}\label{eqn:plm_switching_power}
	2 \rho_\text{t} = (1 + \qo^{-1})^2 + (2y)^2.
\end{equation}
Following the stability analysis of the previous section, (\ref{eqn:plm_order_1_stability}), it is straightforward to show that $\rho_\text{t}$ also corresponds to the power threshold for $\rho_\text{r}$ at which the solution $\rho = 0$ transitions from being stable state to an unstable state.
Hence we might expect $\rho = 0$ for $\rho_\text{r} \leq \rho_\text{r}$ and $\rho = \rho_+$ for $\rho_\text{r} > \rho_\text{r}$.
However, strictly we should also check $\rho = \rho_+$ corresponds to a stable state, as the resonator may simply become unstable above the threshold power.
This requires we demonstrate (\ref{eqn:plm_non_zero_x0_stability_req}) is always true for $\rho = \rho_+$ when $\rho_+ > 0$.
Applying the triangle inequality to the numerator on the left-hand side of (\ref{eqn:plm_non_zero_x0_stability_req}) gives
\begin{equation}\label{eqn:plm_two_circle_stab1}
	| 2 \rho_\text{r} - (1 + \qo^{-1} + \rho_+) \rho_+ |
	\leq 2 \rho_\text{r} + (1 + \qo^{-1} + \rho_+) \rho_+
\end{equation}
where
\begin{equation}\label{eqn:plm_two_circle_stab2}
	 2 \rho_\text{r} + (1 + \qo^{-1} + \rho_+) \rho_+
	 = 4 \rho_\text{r} - (2y)^2 - (1 + \qo^{-1})
	 \sqrt{2 \rho_\text{r} - (2y)^2}.
\end{equation}
The condition $\rho_+ > 0$ can be rearranged to show $(1 + \qo^{-1}) \sqrt{2 \rho_\text{r} - (2y)^2} \geq (1 + \qo^{-1})^2$, which when applied to (\ref{eqn:plm_two_circle_stab1}) and (\ref{eqn:plm_two_circle_stab2}) implies
\begin{equation}\label{eqn:plm_two_circle_stab3}
	| 2 \rho_\text{r} - (1 + \qo^{-1} + \rho_+) \rho_+ |
	\leq 4 \rho_\text{r} - (2y)^2 - (1 + \qo^{-1})^2.
\end{equation}
However, $\rho_+ > 0$ also implies $(2y)^2 - (1 + \qo^{-1})^2 > 2 \rho_r$, so we have succeeded in showing
\begin{equation}\label{eqn:plm_two_circle_stab4}
	| 2 \rho_\text{r} - (1 + \qo^{-1} + \rho_+) \rho_+ |
	\leq 2 \rho_\text{r},
\end{equation}
ensuring (\ref{eqn:plm_non_zero_x0_stability_req}) is true and therefore that $\rho = \rho_+$ is stable state for $\rho_+ > 0$.
Hence, in conclusion we find
\begin{equation}\label{eqn:plm_inv_model_op_point}
	\rho (\rho_\text{r}) = \begin{cases}
		0 & 2 \rho_\text{r} \leq (1 + \qo^{-1})^2 + (2y)^2 \\
		-(1 + \qo^{-1}) + \sqrt{2 \rho_\text{r} - (2y)^2}
		& 2 \rho_\text{r} > (1 + \qo^{-1})^2 + (2y)^2.
	\end{cases}
\end{equation}

\begin{figure}
\centering
\includegraphics{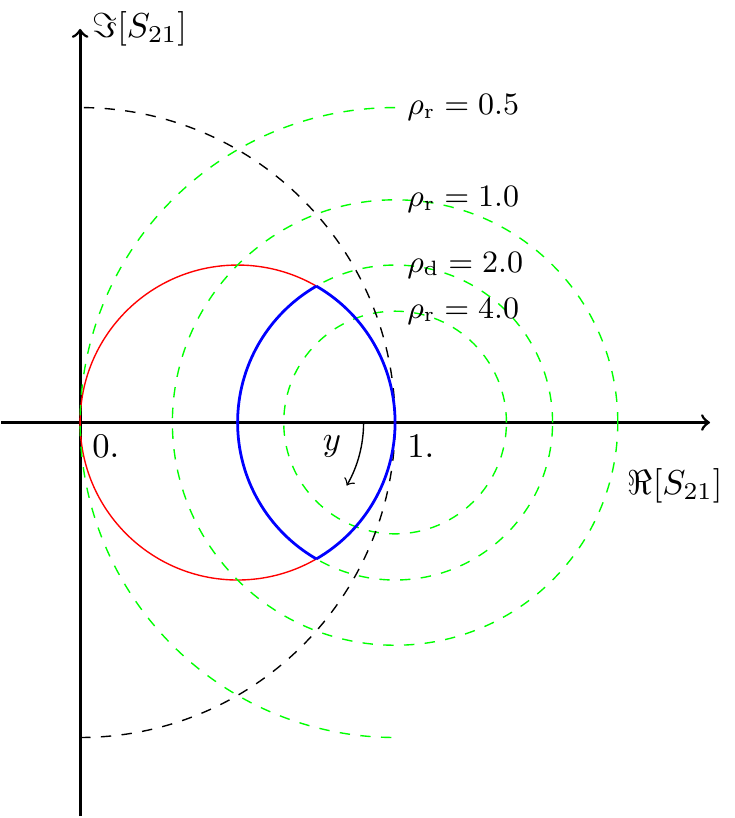}
\caption{\label{fig:plm_two_circle_result} Expected form of the resonance `circle' in the Argand plane for the model described in Section \ref{sec:power_law_model_desc} with $n=d$, or $\Qnl \propto \Pnl^{-1}$.
At high enough powers, the trajectory of $S_{21}$ switches between two distinct circular paths.
}
\end{figure}

(\ref{eqn:plm_inv_model_op_point}) completely determines how the steady-state behaviour of the resonator changes in response to readout power.
Consider the trajectory of $S_{21}$ in the Argand plane as a function of $y$;
(\ref{eqn:plm_inv_model_op_point}) is used to calculate $\qt$, then the result is substituted into (\ref{eqn:mkid_s21}).
After some rearrangement, it can be be shown that $S_{21}$ satisfies
\begin{equation}\label{eqn:plm_s21_below_threshold}
	\left| S_{21} - 1 + \frac{1}{2(1 + \qo^{-1})} \right|
	= \frac{1}{2 (1 + \qo^{-1})},
\end{equation}
below threshold and
\begin{equation}\label{eqn:plm_s21_above_threshold}
	\left| S_{21} - 1 \right|
	= \frac{1}{\sqrt{2 \rho_r}}
\end{equation}
above it.
(\ref{eqn:plm_s21_below_threshold}) and (\ref{eqn:plm_s21_above_threshold}) both describe circular paths in the Argand plane, as illustrated in Figure \ref{fig:plm_two_circle_result}.
The red (solid) circle shows the curve described by (\ref{eqn:plm_s21_below_threshold}), for $\qo^{-1} = 0$ in this case, which is simply the resonance circle that would be traced out by a purely linear device.
The green (dashed) circles show the circles described by (\ref{eqn:plm_s21_above_threshold}) for different values of $\rho_r$.
These are centred on $S_{21} = 1$ and have radius $1 / \sqrt{2 \rho_r}$.
Figure \ref{fig:plm_two_circle_result} can be used to understand the trajectory of $S_{21}$ of the resonator as $y$ is swept from $-\infty$ to $+\infty$.
A device that is below threshold for all $y$, i.e $2 \rho_r \leq (1 + \qo^{-1})^2$, will trace out the red circle clockwise, starting at ending at $S_{21} = 1$.
If $2 \rho_r > (1 + \qo^{-1})^2$, the resonator will be above threshold for at least some values of $y$.
However, it must start below threshold and so $S_{21}$ begins on the green circle, moving clockwise from $S_{21} = 1$.
It will continue along the red (solid) circle until the intersection with the circle for the above threshold solution for $\rho_r$; at this point $2 \rho_r = (1 + \qo^{-1})^2 + (2y)^2$.
A further increase in $y$ moves the device above threshold, so $S_{21}$ starts to move clockwise around the green (dashed) circle.
This gives rise to a sharp point of inflection in the path.
$S_{21}$ will continue along the green (dashed) circle until it intersects the red (solid) circle again, at which point it drops below threshold again and traces the red (solid) path back to $S_{21} = 1$ at $y=\infty$.
The blue (thick solid) line illustrates the overall path for $\rho_r = 2$, illustrating the characteristic distortion pattern associated with the model.

The analysis above can also be linked back to earlier results.
Using (\ref{eqn:plm_s21_above_threshold}) and (\ref{eqn:mkid_s21}), above threshold we have
\begin{equation}\label{eqn:plm_two_circle_s11}
	|S_{11}| = \frac{1}{\sqrt{2 \rho_\text{r}}},
\end{equation}
i.e. $R = |S_{11}|^2$ is maintained at a fixed value by feedback.
This is exactly as was predicted in Section \ref{sec:other_stationary_points}.

\subsection{Experimental observations}\label{sec:plm_exp_data}

\begin{figure}
\centering
\begin{subfigure}{0.48\textwidth}
\centering
\includegraphics[width=\textwidth]{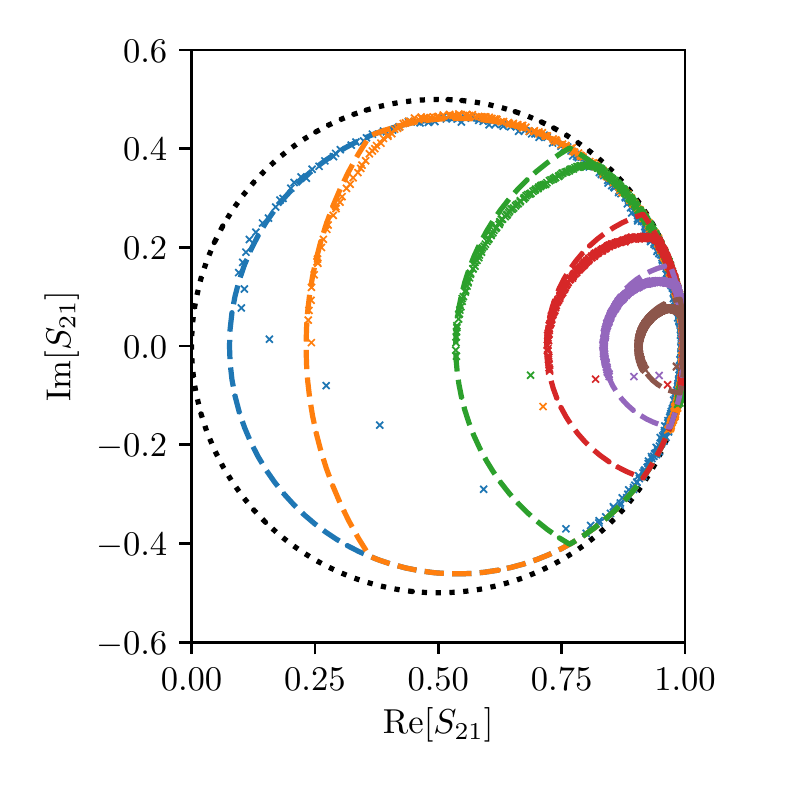}
\caption{\label{eqn:plm_two_circle_exp_s21}
$S_{21}$ in the Argand plane as measured on a downward frequency sweep for VNA power levels -80\,dBm (blue), -75\,dBm (orange), -70\,dBm (green), -65\,dBm (red), -60\,dBm (purple) and -55\,dBm (brown).
The crosses indicate the measured data points and the dashed line of matching colour the model fit to the data, as described in the text.
The dashed black line is a circle of unit radius centered on $S_{21} = 0.5$.
}
\end{subfigure}
\quad
\begin{subfigure}{0.48\textwidth}
\centering
\includegraphics[width=\textwidth]{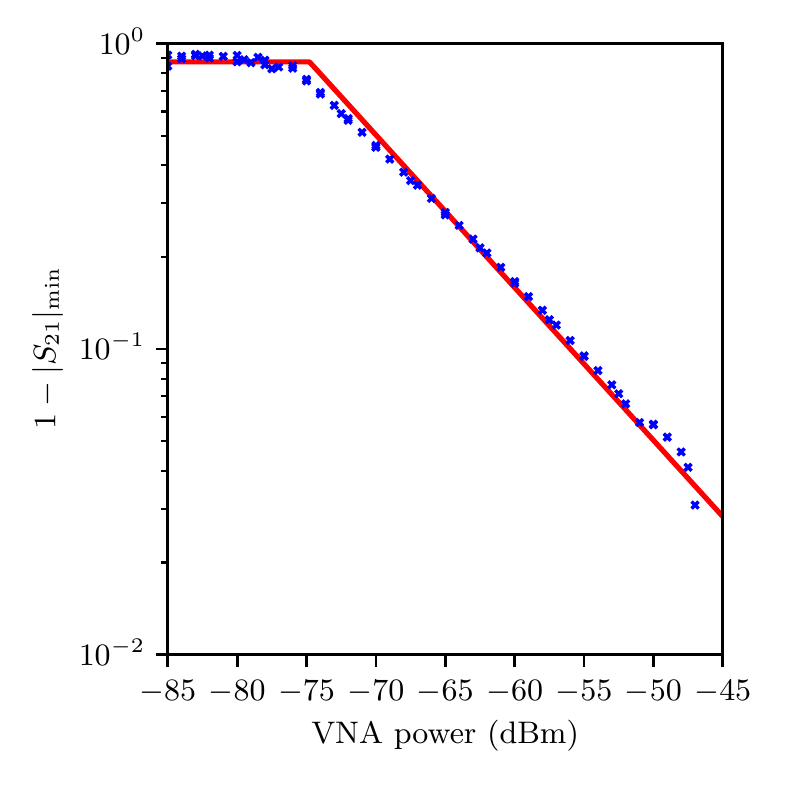}
\caption{\label{eqn:plm_two_circle_exp_depth}
$1 - |S_{21}|_\text{min}$ versus applied readout power as measured at the VNA output port.
The blue crosses are the measured data points, while the red line shows the line of best fit resulting from a least-squares fit using (\ref{eqn:plm_two_circle_depth_vs_power}).
\vspace{4\baselineskip}
}
\end{subfigure}
\caption{\label{fig:plm_two_circle_experimental_data}
Experimental data showing behaviour similar to that predicted by the model of Section \ref{sec:nlm_exp_equals_one}.
}
\end{figure}

We have observed the remarkable behaviour described in Section \ref{sec:nlm_exp_equals_one} in many resonators.
One such device is the resonator with the higher $\Qc$ out of the two NbN devices described previously, in Section \ref{sec:qp_comparison_with_exp}.

Figure \ref{eqn:plm_two_circle_exp_s21} shows $S_{21}$ of this device in the Argand plane as measured on a downward frequency sweep for different readout power levels.
The mappings of the different curves to readout power are given in the figure caption.
In each case the crosses show the experimental data and the dashed line of matching colour a fit of the model from Section \ref{sec:nlm_exp_equals_one}.
The large discontinuities in the data in the lower half of the plot are the result of the presence of a simultaneous reactive nonlinearity, which results in switching.
As can be seen, the model and data are generally in very good agreement.
The only place they differ is at the threshold where $S_{21}$ switches between circles; in the data this transition is softer than the model predicts.
By using the full power law model we found that this behaviour can be reproduced by using a value of $n/d$ close to but slightly less than one.

Figure \ref{eqn:plm_two_circle_exp_depth} is a plot of measured resonance depth (blue crosses) as a function of readout power.
Resonance depth is taken here to be the difference between the transmission far off resonance and the minimum transmission in resonance, i.e. $1 - |S_{21}|_\text{min}$ for a de-embedded device.
Given the model, we would expect
\begin{equation}\label{eqn:plm_two_circle_depth_vs_power}
	1 - |S_{21}|_\text{min}
	= \qt = \begin{cases}
		1 & 2 \rho_\text{r} \leq (1 + \qo^{-1})^2 + (2 y)^2 \\
		\frac{1}{1 + \sqrt{2 \rho_\text{r}}}
		& 2 \rho_\text{r} > (1 + \qo^{-1})^2 + (2 y)^2.
	\end{cases}
\end{equation}
At a fit of this model to the data, allowing $\Pcpl$ to vary, is shown by the blue line in Figure \ref{eqn:plm_two_circle_exp_depth}.
The agreement between model and data is again very good.
However, if anything, the gradient of the data is slightly shallower than the model would predict.
This would suggest a value of $n/d$ slightly less than one, which is consistent with the observations of the shape of the resonance `circles'.
As described in Section \ref{sec:qp_comparison_with_exp}, there is strong evidence the underlying physical mechanism is quasiparticle heating in this case.
However, it has also been shown that superconducting weak links can play a role in nonlinear behaviour in NbN resonators\,\cite{abdo2006observation}.

\subsection{Power law exponent greater than one}\label{sec:nlm_exp_greater_than_one}

\begin{figure}
\centering
\begin{subfigure}{0.45\textwidth}
\includegraphics[width=\textwidth]{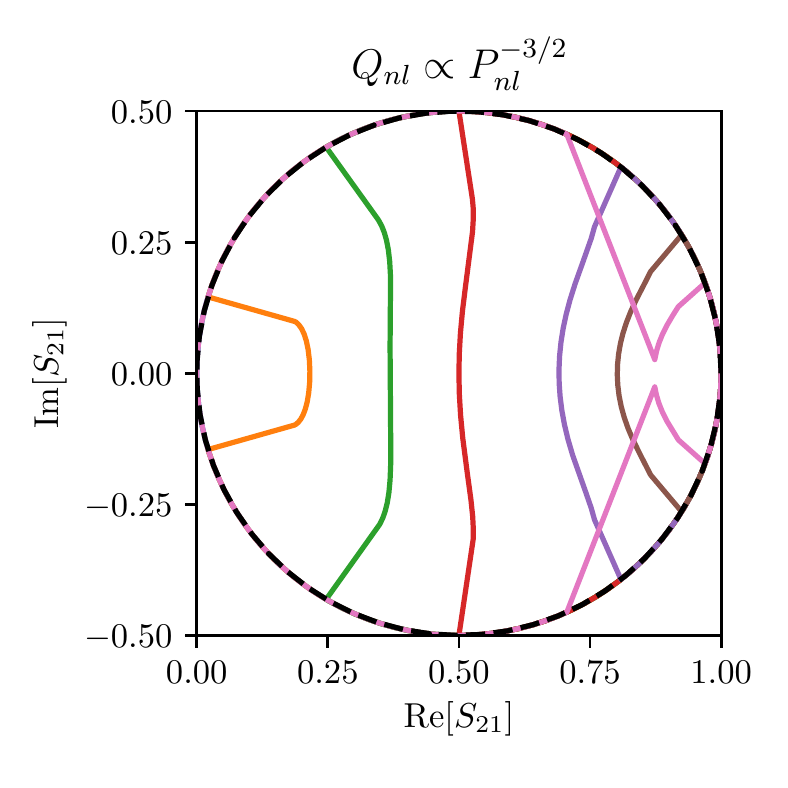}
\caption{\label{fig:plm_argand_circle_plots_n_3_d_2}
Resonance circles are shown for $\rho_\text{r} = $ 1.0 (under dashed black line), 1.25 (orange), 1.5 (green), 2 (red), 4 (purple), 8 (brown) and 16 (pink).
}
\end{subfigure}
\quad
\begin{subfigure}{0.45\textwidth}
\includegraphics[width=\textwidth]{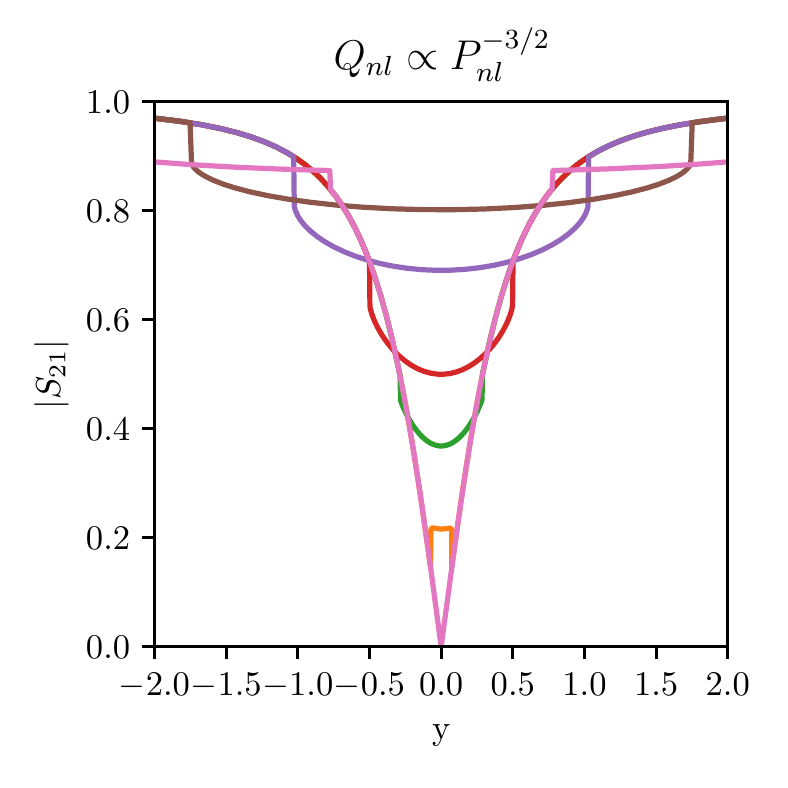}
\caption{\label{fig:plm_amp_plots_n_3_d_2}
$|S_{21}|$ as a function of $y$ for $\rho_\text{r} = $ 1.25 (orange), 1.5 (green), 2 (red), 4 (purple), 8 (brown) and 16 (pink).
Note the discontinuities in the amplitude in the wings of the resonance feature.
}
\end{subfigure}
\caption{\label{fig:plm_exponent_greater_than_one}
Example of the behaviour of the model of described in Section \ref{sec:power_law_model_desc} when $n=3$, $d=2$ and $\qo = 10^8$, i.e. when the power law exponent is greater than one.
}
\end{figure}

The behaviour for $n / d > 1$ is significantly different and much more complicated than the other cases, as illustrated by the plots in Figure \ref{fig:plm_exponent_greater_than_one}.
These plots show a set of simulated curves for different $\rho_\text{r}$ for the case $n=3$ and $d=2$.
Figure \ref{fig:plm_argand_circle_plots_n_3_d_2} shows the measured resonance curves in the Argand plane, while Figure \ref{fig:plm_amp_plots_n_3_d_2} shows the measured amplitude of $S_{21}$ as a function of the applied detuning.  

Below $\rho_\text{r} \approx 1.1$, $\rho = 0$ is the only solution.
For $\rho_\text{r} = 1.25$ (the orange line in Figure \ref{fig:plm_argand_circle_plots_n_3_d_2}), we see the formation of a feature near $y=0$.
When this feature is viewed on a plot of amplitude versus detuning, it appears as a small peak in $S_{21}$ at the bottom of the resonance trough (Figure  \ref{fig:plm_amp_plots_n_3_d_2}).
As $\rho_\text{r}$ is increased further this feature opens out and folds back on itself, leading to shapes reminiscent of those for the case $n/d \leq 1$, e.g. the purple and brown curves.
However, when $\rho_\text{r}$ is further increased we see a surprising new feature arise where near $y=0$ where the device switches back to the state $\rho = 0$ in the region where dissipation should be strongest.
This suggests there is a high power state at which the dissipative state can effectively switch itself off; the rate of increase in dissipation with $\rho_\text{r}$ is sufficient that the dissipated power actually begins to fall with increased $\rho_\text{r}$, so the dissipation cannot sustain itself.

What is not clear from Figure \ref{fig:plm_argand_circle_plots_n_3_d_2} is that the trajectory of $S_{21}$ in the Argand plane also becomes discontinuous.
This is better illustrated by Figure \ref{fig:plm_amp_plots_n_3_d_2}, which shows $|S_{21}|$ as a function of $y$ for $\rho_\text{r} = 2$, 4, 8 and 16.
As can be seen, there are now step discontinuities in $|S_{21}|$ in the wings of the resonance feature.
These occur where $S_{21}$ departs from the circle for $\rho = 0$ in the Argand plane.

What may complicate the observation of such behaviour in practice is the fact the state $\rho = 0$ is also always stable for $n / d > 0$.
As discussed before, which state the device ends up in will depend on how the device has been prepared, e.g. is $y$ or power being swept?
Without further detailed analysis it is not possible to say what method, if one exists, is needed to see the unusual behaviour shown.

Similar step discontinuities to those shown in Figure \ref{fig:plm_argand_circle_plots_n_3_d_2} has been observed by Abdo\,\cite{abdo2006observation} in a set of NbN resonators.
In addition, they observe hysteresis around these steps with sweep direction.
This latter behaviour can be explained by the resonator switching from a state with $\rho > 0$ to the one with $\rho = 0$ at the first transition point, then remaining in this state as it passes through the location of the second discontinuity.
They also see the on-resonance transmission initially increase with increasing readout power, then jumping suddenly to a fixed, higher, value; this is consistent with the behaviour predicted by Figure \ref{fig:plm_exponent_greater_than_one} if the device were a transmission resonator.
They attribute this behaviour to either weak-link formation in the NbN grain structure or, alternatively, the formation of localised hot spots.

\section{Simultaneous action of several mechanisms}\label{sec:multiple_mechanisms}

We have considered each nonlinear process acting in isolation, but in some cases, it is the interaction between different processes that determines behaviour.
As an example, consider a resonator limited by TLS loss.
The results of Section \ref{sec:tls} when taken alone suggest that the quality factor can be improved by increasing the readout power so as to saturate the TLS.
However, at some point as the readout power is increased, quasiparticle heating may become significant, resulting in the quality factor decreasing as the power is increased further.
The maximum achievable quality factor is determined by the interplay of the two processes, and their relative characteristic power scales.
This `sweet spot' is the operating regime often chosen for best device performance.
In extreme cases, we have observed that quasiparticle heating can prevent TLS saturation, and so the quality factor only decreases as power is applied.

Given the importance of these effects, it is valuable to consider how the models presented can be modified to include interactions.
The procedure is conceptually straightforward, but computationally involved.
A single variable fully characterises the `state' of the nonlinearity for each process considered: $U$ for the reactive non-linearity and TLS loss, $\nqp$ for quasiparticle heating, and $\Pnl$ for a general physically unidentified nonlinearity.
Further, for a particular set of readout conditions the value of this state parameter is found by solving a single equation, often an equilibrium or self-consistency condition:
(\ref{eqn:tls_master_eqn}), (\ref{eqn:qp_master_equation}) and (\ref{eqn:plm_qnl}).
It is therefore possible to model several processes acting together by solving these equations simultaneously, replacing the $\Qo$ term in the individual models by the contributions from other processes.
We have developed a convenient conceptual framework for structuring these calculations and easily including new processes.
However, space precludes a full description of the method and an exploration of the rich set of behaviours that results.
Instead they will be detailed in a companion publication\,\cite{skyrme2020understanding}.

\section{Extracting behaviour from data}\label{sec:measurement_scheme}

Finally, we indicate how key parametric information can be determined easily from experimental data.
It is normally straightforward to record a set of swept-frequency resonance curves at different readout power levels using a VNA or homodyne readout system.
The difficulty lies in extracting the underlying nonlinear behaviour when the resonance curves become distorted.
In other words, distorted resonance curves are merely manifestations of the change in the resonance frequency and Q of the underlying simple Lorentzian resonance changing as the readout frequency and power are varied.
In principle, we could fit a full nonlinear model of the type described in Section \ref{sec:tls}--\ref{sec:multiple_mechanisms} and obtain the associated physical parameters, but to do so we need to know the expected nonlinear behaviour in advance.
Additionally, as the model becomes more complex so does the fitting process.
Section \ref{sec:distortion}, however, motivates a different approach.

The aim is to directly extract the quality factor and resonant frequency at zero realised detuning, for different readout power levels.
To do so, we must ensure that the swept-frequency measurements pass through the point of zero realised detuning.
This is discussed in Section \ref{sec:accessing_zero_detuning}, and the process is  normally straightforward; for example, if the resonant frequency is known to decrease with applied power, the frequency must be swept downwards when the resonance curves are measured.
Next we must identify the point of zero realised detuning in each resonance curve.
The rules derived in Section \ref{sec:zero_detuning} can be used to do so: this is as simple as finding the extrema in the transmission gain or point of zero phase shift.
Finally, having located the point, the resonant frequency follows from the readout frequency, and the quality factors from the measured S-parameter using (\ref{eqn:mkid_s11})--(\ref{eqn:mkid_s21_s}).
This process is repeated to give the key parameters as a function of applied power. 

This method has several attractive features.
First, the data and processing needed are straightforward.
Second, it is applicable to highly distorted curves, and can therefore be used over wide power ranges.
In other words, it is still possible to extract mathematically meaningful, and physically well-defined, resonance frequencies and quality factors, even though the measured resonance curves switch hysteretically, and bear no resemblance to simple Lorentzians.
Third, by definition we know the realised detuning at which the parameters were obtained, and this makes it straightforward to convert the applied readout power into the quantities that control the nonlinear behaviour.

As an example, consider a resonator exhibiting a mixture of reactive and dissipative nonlinear behaviour.
Assume that the reactive nonlinearity results in Duffing-like behaviour with an increasingly negative frequency shift at high readout powers.
To apply our parameter-extraction scheme a set of swept-frequency resonant curves would be recorded at different readout power levels, being careful to sweep the readout frequency \emph{downwards} in each measurement, which is in the opposite direction to the usual VNA settings.
The recorded data would then be processed by first removing any experimental artefacts, such as gain- and phase slopes.
The maximum in transmission gain of each resonance curve would be located, checked against phase, and used to calculate values of $\fres$ and $\Qt / \Qc$ at the corresponding readout power and $x=0$ via (\ref{eqn:fres_at_zero_detuning}) and (\ref{eqn:qi_at_zero_detuning}).
The data shown in Figures \ref{fig:example_data_qi_vs_pvna} and \ref{fig:plm_two_circle_experimental_data} was taken in this manner.

\section{Conclusions}\label{sec:conclusions}

Superconducting thin-film resonators are used extensively in many applications.
They can take a variety of physical forms, and can be fabricated using a wide range of materials, including proximitised superconducting multilayers.
From a device perspective, it is usually assumed that the resonator alone acts as an a near-ideal linear device, exhibiting a perfect response in the form of a Lorentizian notch or peak.
In reality this simple behaviour is rarely seen, and non-linear behaviour becomes apparent when the readout power is increased to optimise some aspect of overall device performance.

We have discussed how reactive and dissipative non-linearities can, and do, change the intrinsic response of thin-film resonators considerably, leading to complex behaviour that can mask or degrade the primary device-operation being sought.
At its most minor, resonance curve distortion can indicate heating, which may increase the noise generated by the device; at its most significant, resonance curve distortion can be associated with hysteretic switching between different stable states, and the operating point can depend on the order in which the external parameters are changed.

We have shown that most, if not all, of the complex phenomena commonly seen in experiments can be described by a model in which the underlying resonance is a single-pole Lorentizian, but whose centre frequency and quality factor change depending on the energy stored in the resonator and/or the power dissipated in various physical processes.
What is seen experimentally are samples taken from an ideal resonance curve that is moving and changing width as external parameters, such as readout frequency and power, are swept.
According to this model, it is perfectly proper to refer to, and to measure, the Q of the underlying resonance, even though the swept frequency curves appear highly distorted and perhaps hysteretic.
Indeed, there is a great deal of information contained in the parametric dependence of the $Q$ of the underlying resonance, not just in the resonant frequency.
In those cases where the resonance curve is highly distorted, the shape of the trajectory in the Argand plane gives valuable insights into the physical processes present.

Kinetic inductance is an example of a reactive nonlinearity, which leads to a shift in the resonance frequency, and eventually hysteretic switching, but the trajectory in the complex plane remains circular.
The point of zero detuning is important, and can still be found from zero crossings and stationary points in the transmission and reflection amplitudes, as for a linear device.
Two Level Systems in oxides primarily introduce a dissipative nonlinearity.
We have described a fixed point method for calculating measured resonance curves, and shown how the trajectory in the Argand plane takes on a characteristic `tear drop' shape.
We have also shown that TLSs cannot produce hysteresis, but they lead to a phenomenon, seen experimentally, where an apparently absent resonance suddenly switches on as the readout power is increased.
Quasiparticle heating leads to a completely different kind of dissipative nonlinearity.
Sub-gap readout photons change the energy distribution and number density of quasiparticles, which themselves change the dissipation factor.
We have presented a model based of the Rothwarf Taylor equations that gives a simple expression for the internal quality factor as a function of readout power.
This formulation leads to a scheme in which resonator dynamics is described by a quartic equation, and we discussed the stabilities of the roots of this equation under different coupling conditions.
We find different behaviours in the undercoupled and overcoupled cases, due to the existence of negative and positive feedback respectively in the quasiparticle generation process.
Crucially, the trajectory in the complex plane takes on a highly characteristic two-part piecewise circular form.
In this case, the points of zero detuning can be identified directly, and the quality factor of the underlying resonance found.
Finally, we introduced a generic power law model, where the internal quality factor depends on the dissipated power raised to the power of a rational number.
This generic model captures the key features of specific dissipative non-linearities, but additionally leads to insights into how general dissipative processes create characteristic forms of behaviour in the Argand plane.
We have found these insights to be highly valuable when interpreting the rich variety of behaviour seen experimentally in different kinds of device.

\bibliographystyle{unsrtnat}
\newcommand*{\doi}[1]{\href{http://dx.doi.org/#1}{doi: #1}}
\bibliography{references}

\appendixpage
\begin{appendix}

\section{Additional results from Swenson's model}\label{sec:swenson_additional}

\subsection{Point of onset of hysteresis}\label{sec:swenson_bifurcation_point}

For notational convenience define $z = y / (1 + \eta)$ and $z_0 = y_0 / (1 + \eta)$.
Then we can rewrite (\ref{eqn:swenson_y_equation}) as an equation defining the roots of the cubic polynomial
\begin{equation}\label{eqn:swenson_poly_in_z}
	f(z) = z^3 - z_0 z^2 + z / 4 - (z_0 + a) / 4.
\end{equation}
For hysteresis to occur, $f(z)$ must have three real roots.
A sufficient condition to ensure this is that $f(z)$ has two stationary points for real $z$ and that $f(z)$ differ in sign at these points.
Factor $f(z)$ as
\begin{equation}\label{eqn:factored_swenson_poly_in_z}
\begin{aligned}
	f(z)
%	&= z^3 - z_0 z^2 + z / 4 -  (z_0 + a) / 4 \\
%	&= (z - z_0 / 3)^3 - (-z_0 z^2 + z_0^2 z / 3 - z_0^3 / 27)
%		- z_0 z^2 + z / 4 -  (z_0 + a) / 4 \\
%	&= (z - z_0 / 3)^3 + z_0 z^2 - z_0^2 z / 3 + z_0^3 / 27
%		- z_0 z^2 + z / 4 -  (z_0 + a) / 4 \\
%	&= (z - z_0 / 3)^3 - (z_0^2 / 3 - 1/ 4) z
%		+ z_0^3 / 27 - (z_0 + a) / 4 \\
%	&= (z - z_0 / 3)^3 - (z_0^2 / 3 - 1/ 4) (z - z_0 / 3)
%		- (z_0^2 / 3 - 1/ 4) z_0 / 3 + z_0^3 / 27 - (z_0 + a) / 4 \\
%	&= (z - z_0 / 3)^3 - (z_0^2 / 3 - 1/ 4) (z - z_0 / 3)
%		- z_0^3 / 9 + z_0 / 12 + z_0^3 / 27 - (z_0 + a) / 4 \\
%	&= (z - z_0 / 3)^3 - (z_0^2 / 3 - 1/ 4) (z - z_0 / 3)
%		- (2 z_0^3 / 27 + z_0 / 6 +  a / 4) \\
%	&= (z - z_0 / 3)^3 - (z_0^2 - 3 / 4) (z - z_0 / 3) / 3
%		- (2 z_0^3 / 27 + z_0 / 6 +  a / 4) \\
%	&= (z - z_0 / 3)^3 - (z_0^2 - 3 / 4) (z - z_0 / 3) / 3
%		- (2 z_0^3 / 9 + z_0 / 2 +  3 a / 4) / 3 \\
	&= (z - z_0 / 3)^3 - c (z - z_0 / 3) - d,
\end{aligned}
\end{equation}
where $c = (z_0^2 - 3/4) / 3$ and $d = (2 z_0^3 / 9 + z_0 / 2 + 3a / 4) / 3$.
In this form it is straightforward to see that there are two stationary points only when
\begin{equation}\label{eqn:condition_on_z0}
	|z_0| > \sqrt{3}/2,
\end{equation}
that they occur at $z = z_\pm$ for
\begin{equation}\label{eqn:stationary_z}
	z_\pm - \frac{z_0}{3} = \frac{1}{3} \sqrt{z_0^2 - \frac{3}{4}},
\end{equation}
and that
\begin{equation}\label{eqn:f_at_stationary_z}
	f(z_\pm) = \mp \frac{2c}{3} \sqrt{\frac{c}{3}} - d.
\end{equation}
Hence we require
\begin{equation}\label{eqn:condition_on_c_and_d}
	\frac{2c}{3} \sqrt{\frac{c}{3}} > |d|
\end{equation}
for the signs of $f(z_+)$ and $f(z_-)$ to differ.

From the analysis earlier in the paper we know the hysteretic regime occurs where $z_0 < 0$.
Let $z_0 = -\sqrt{3} / 2 - \delta$ for $0 < \delta \ll \sqrt{3} / 2$, in which case
\begin{equation}\label{eqn:c_approx}
	\frac{2c}{3} \sqrt{\frac{c}{3}}
	\approx 2 \left( \frac{\delta}{3 \sqrt{3}} \right)^{3/2}
\end{equation}
and
\begin{equation}\label{eqn:d_approx}
	d \approx -\frac{\sqrt{3}}{9} + \frac{a}{4} - \frac{\delta}{3}.
\end{equation}
(\ref{eqn:condition_on_c_and_d}) is then satisfied when
\begin{equation}\label{eqn:a_condition}
	a > \frac{4 \sqrt{3}}{9} + \frac{4 \delta}{3}
	-  \left( \frac{4 \delta}{3 \sqrt{3}} \right)^{3/2}.
\end{equation}
The two terms in $\delta$ in (\ref{eqn:a_condition}) always sum to a positive number for $\delta < 1$, so the overall threshold for switching is $a > 4 \sqrt{3} / 9$.

\subsection{Location of the switching point on a downward sweep}\label{sec:loc_switching_points}

(\ref{eqn:up_and_un_equation}) can be rearranged to yield the following iterative sequence for $u_-$:
\begin{equation}\label{eqn:up_and_un_iteration}
	u^{(n+1)}_- = -\frac{1}{8a} (1 + 4 \{ u^{(n)}_-\}^2)^2.
\end{equation}
Either by iteration for a few terms or by examination of this sequence, it can be seen that in general
\begin{equation}\label{eqn:un_approx}
	u_- = -\frac{1}{8a} + O \left( \frac{1}{a^3} \right).
\end{equation}
This result can then be substituted into (\ref{eqn:swenson_y_equation}) to find the value $y_-$ of $y_0$ at which the resonator is expected to switch states on a downward sweep, yielding
\begin{equation}\label{eqn:yn_approx}
	y_- \approx -(1 + \eta) a + O \left( \frac{1}{a} \right),
\end{equation}
in the limit $a \gg 1$.
Since both $y_-$ and $\eta$ can be easily measured in such a sweep, (\ref{eqn:yn_approx}) provides a convenient way of estimating $a$ experimentally, either as a starting point for a fit or for inferring $\Uckin$.
This approach is slightly more straightforward than that proposed in Swenson\,\cite{swenson2013operation}, which involves identifying the onset of bifurcation ($a \approx 0.8$).

\section{Proofs relating to the TLS model}\label{sec:tls_proofs}

\subsection{Proof that solution of (\ref{eqn:tls_fixed_point_problem}) exists}\label{sec:existance_of_tls_solution}

Let $I$ denote the interval $[0, 1]$, which corresponds to the range of values of $\alpha$, where $\alpha = f(\alpha)$.
We will use square brackets to denote an interval limit that includes the end point and curved brackets to indicate a limit that excludes the end point.
For example, the interval $[a, b]$ of $x$ corresponds to $a \leq x \leq b$ and $[a, b)$ to $a \leq x < b$.
Given definitions (\ref{eqn:tls_def_f}) and (\ref{eqn:tls_def_chi}), it is straightforward to show that for the problem in hand
\begin{equation}\label{eqn:tls_deriv_f_wrt_alpha}
	\frac{df}{d\alpha} = \frac{r \ytls \chi^2}{[(1 - r \alpha)^2 + \chi \ytls]^{3/2}}.
\end{equation}
We know $\ytls \geq 0$, $0 \leq r \leq 1$ and $0 \leq \chi \leq 1$, so $df/d\alpha \geq 0$ for any real $\alpha$.
It follows that $f(\alpha)$ is increasing function on $I$, with the consequence $f(0) \leq f(\alpha) \leq f(1)$ for $\alpha \in I$.
These limits are explicitly
\begin{equation}\label{eqn:tls_f0}
	f(0) = 1 - \frac{1}{\sqrt{1 + \chi \ytls}}
\end{equation}
and
\begin{equation}\label{eqn:tls_f1}
	f(1) = 1 - \frac{1}{\sqrt{1 + \chi \ytls / (1 - r)^2}}.
\end{equation}
The conditions on $\chi$ and $\ytls$ ensure $\chi \ytls \geq 0$, so we have $0 \leq f(0), f(1) \leq 1$.
Therefore, $f(\alpha) \in [0, 1]$ for all $\alpha \in [0, 1]$.

The last statement is sufficient to ensure the existence of at least one solution of $f(\alpha) = \alpha$, i.e. (\ref{eqn:tls_fixed_point_problem}), with $\alpha \in I$, via the one-dimensional form of Brouwer's fixed-point theorem.
The proof is as follows.
Consider a new continuous function $h(\alpha) = f(\alpha) - \alpha$.
If $f(0) = 0$ or $f(1) = 1$, then we trivially have a solution to (\ref{eqn:tls_fixed_point_problem}).
If not, we know $f(0) > 0$ and $f(1) < 0$ and this implies $h(0) > 0$ and $h(1) < 0$ respectively.
It follows by the intermediate-value theorem\,\cite{jeffreys1956methods} that $h$ must have at least one root in $I$, with the existence of this root implying (\ref{eqn:tls_fixed_point_problem}) is satisfied.

\subsection{Proof of convergence of (\ref{eqn:tls_iterative_sequence}) and physical uniqueness of solution for \texorpdfstring{$x = 0$}{x = 0}}\label{sec:tls_solution_zero_x}

We will make use of the following fixed-point theorem: if a function $g(x)$ maps an interval $I$ into itself and $|dg/dx| < 1$ for $x \in 1$, then $g(x)$ has a unique fixed point $x=f(x)$ that is the limit $n \rightarrow \infty$ of the sequence $x_n = g(x_{n-1})$ for $x_0 \in I$.
This is the one-dimensional form of Banach's fixed-point theorem.
In Section \ref{sec:existance_of_tls_solution} we showed $f(\alpha)$ maps the interval $I$ into itself, so to prove (\ref{eqn:tls_iterative_sequence}) converges we only need to consider the conditions on the derivative.

If $x=0$, then $\chi = 1$ for all $\alpha$.
With $\chi = 1$ in (\ref{eqn:tls_deriv_f_wrt_alpha}), we can define three cases to cover all possible physical situations.
Case 1 is where $\ytls > r^2$, so $df/d\alpha < 1$ for all $\alpha$.
If instead $\ytls \leq r^2$, it is straightforward to show that $df/d\alpha < 1$ if $\alpha$ is less than
\begin{equation}\label{eqn:tls_alpha_star}
	\alpha_* = \frac{1}{r} \left[ 1 - \sqrt{\ytls^{2/3} - \ytls} \right].
\end{equation}
Cases 2 and 3 are where $\alpha_* > 1$ and $\alpha_* < 1$ respectively.
In Cases 1 and 2, $f(\alpha)$ satisfies the fixed-point theorem over the whole of $I$.
Consequently, (\ref{eqn:tls_iterative_sequence}) converges to the unique physical solution for any starting value of $\alpha$ in $I$.

In Case 3, $f(\alpha)$ no longer satisfies the condition on the derivative over the whole of $I$.
However, remembering that $r < 1$ (by definition) and that Case 3 requires $\ytls > r^2$, it is simple to prove
\begin{equation}\label{eqn:tls_fstar}
	f(\alpha_*) = 1 - \frac{\sqrt{\ytls^{2/3} - \ytls}}{(r \ytls)^{1/3}}
	< \alpha_*.
\end{equation}
Consequently, $f(\alpha)$ satisfies the conditions of the fixed point theorem on the reduced interval $[0, \alpha_*)$, so (\ref{eqn:tls_iterative_sequence}) will converge to a single physical solution for suitable starting point.
However, we cannot yet say that the solution found is this manner is the only physically possible one; to do so we must show there are no other fixed-points in the interval $[\alpha_*, 1]$.
The proof of the latter statement is as follows.
Consider again the function $h(\alpha)$ introduced in Section \ref{sec:existance_of_tls_solution}.
In Case 3, $dh/d\alpha \geq 1$ over $[\alpha_*, 1]$, making $h(\alpha)$ an increasing function over the same interval.
Because we know that $h(\alpha_*), h(1) < 0$ from (\ref{eqn:tls_fstar}) and Section \ref{sec:existance_of_tls_solution}, we can then use the fact $h(\alpha)$ is increasing to show $h(\alpha) < 0$ over $[\alpha_*, 1]$.
The latter statement precludes the existence of a fixed-point of $f(\alpha)$ in $[\alpha_*, 1]$; the fixed-point in $[0, \alpha_*)$ is therefore the only physical solution.

\subsection{Proof of convergence of (\ref{eqn:tls_iterative_sequence}) and physical uniqueness of solution for \texorpdfstring{$x\neq 0$}{nonzero x}}\label{sec:tls_solution_non_zero_x}

When $x \neq 0$ the full functional dependence of $\chi$ on $\alpha$, as given by (\ref{eqn:tls_def_chi}), must be taken into account.
If $\chi$ is treated as an independent variable in (\ref{eqn:tls_def_f}) and (\ref{eqn:tls_deriv_f_wrt_alpha}), then it can be shown that both $f(\alpha)$ and $df/d\alpha$ are increasing functions of $\chi$ for $\chi \geq 0$ and $\alpha \in [0, 1]$.
However, $\chi$ is actually a decreasing function of $\alpha$ on the same interval when $x \neq 0$.
It follows that both $f(\alpha, x) \leq f(\alpha, x=0)$ and $df(\alpha, x)/d\alpha \leq df(\alpha, x=0)/d\alpha$ on this interval.

These last two inequalities mean the proof of Section \ref{sec:tls_solution_zero_x} in Cases 1 and 2 and the first part of Case 3 extend trivially to $x \neq 0$.
The proof of the second part of Case 3 follows straightforwardly; if in some interval $f(\alpha, x=0)$ is bounded above by $\alpha$ and $f(\alpha, x)$ is bounded above by $f(\alpha, x=0)$, then it is not possible for $f(\alpha, x)$ to intersect $\alpha$.
The results of Section \ref{sec:tls_solution_zero_x} therefore also hold when $x \neq 0$.

There is also an important physical consequence to this result.
By showing that there is a only a single physical solution of (\ref{eqn:tls_fixed_point_problem}) for real $x$ and $\ytls$ with $\ytls \geq 0$, we have ruled out the possibility of hysteretic behaviour when TLS response is the only source of nonlinearity.

\section{Dependence of the quasiparticle quality factor on quasiparticle density}\label{sec:qqp_as_a_function_of_n}

Let $\sigma = \sigma_1 - i \sigma_2$ denote the bulk conductivity of a superconductor, with $\sigma_1$ and $\sigma_2$ both real.
Gao\,\cite{gao2008equivalence,gao2008physics} has shown that in the low-frequency ($h \nu \ll 2 \Delta$), low-temperature ($T / \Tc < 1$), regime in which superconducting resonators are employed, the Mattis-Bardeen\,\cite{mattis1958theory} equations for $\sigma$ can be approximated by
\begin{equation}\label{eqn:mb_approx_sigma1}
	\frac{\sigma_1}{\sigma_\text{n}}
	= \frac{2 \Delta_0}{h \nu}
	\frac{\nqp}{N_0 \sqrt{2 \pi k_\text{b} T \Delta_0}}
	\sinh \left( \frac{h \nu}{2 k_\text{b} T} \right)
	K_0 \left( \frac{h \nu}{2 k_\text{b} T} \right)
\end{equation}
and
\begin{equation}\label{eqn:mb_approx_sigma2}
	\frac{\sigma_2}{\sigma_\text{n}}
	= \frac{\pi \Delta}{h \nu}
	\left[ 1 - \frac{\nqp}{2 N_0 \Delta_0}
	\left( 1 - \sqrt{\frac{}{}}
	e^{-h \nu / 2 k_\text{b} T}
	I_0 \left( \frac{h \nu}{2 k_\text{b} T} \right)
	\right) \right].
\end{equation}
Here $\sigma_\text{n}$ is the normal state conductivity and $\Delta_0$ the superconducting gap energy at absolute zero.
These results can understood physically in terms of a two-fluid model.
In the regime considered the dominant charge carriers are the Cooper pairs, which move without scattering and hence do not contribute the real part of the conductivity, $\sigma_1$.
Instead, their inertia manifests itself as an inductance like term as described by $\sigma_2$ (kinetic inductance).
However, some fraction of the Cooper pairs are broken into quasiparticles, either by thermal processes or by external forcing.
This loss of Cooper pairs reduces the inductive response, as described by the second term in (\ref{eqn:mb_approx_sigma2}).
In addition, the quasiparticles behave electrically approximately like normal state Drude model electrons, leading to a resistance contribution proportional to $\nqp$: (\ref{eqn:mb_approx_sigma1}).
Although it is not immediately obvious from (\ref{eqn:mb_approx_sigma1}) and (\ref{eqn:mb_approx_sigma2}), $\sigma_2 \gg \sigma_1$ in this regime.
Further, we can usually make the further approximation $\sigma_2 / \sigma_\text{n} \approx \pi \Delta_0 / h \nu$.

We must now link $Q$ with $\sigma$.
In the case of a lumped element device, this is relatively straightforward.
This is because the superconductor film is normally used in a regime where it is electrically thin and the contribution from geometric reactance is small, so it can be approximated as an impedance $Z$ given by
\begin{equation}\label{eqn:lumped_element_impedance}
	\frac{1}{Z} = \frac{\sigma t}{N_\text{sq}},
\end{equation}
where $t$ is the film thickness and $N_\text{sq}$ is the length of the superconducting trace expressed in squares.
$Z$ constitutes the parallel inductance $L$ and resistance $R$ in (b) of Figure \ref{fig:different_resonator_geometries}.
Using the normal result for the quality factor of a parallel tank circuit, we find
\begin{equation}\label{eqn:q_parallel_lcr}
	Q^{-1} = \frac{2 \pi \nu L}{R} = \frac{\sigma_1}{\sigma_2}.
\end{equation}
Making use of (\ref{eqn:mb_approx_sigma2}) we then have $Q^{-1} \propto \nqp$,
as assumed in (\ref{eqn:qp_qi_from_nqp}).

In the case of a transmission line resonator of length $l$, if $\gamma$ is the complex propagation constant of waves on the line then it can be shown that
\begin{equation}\label{eqn:tl_q_from_gamma}
	\Qi^{-1} \propto \Re[\gamma] l.
\end{equation}
Strictly this expression accounts for both Ohmic and dielectric losses; in what follows we will assume there are only Ohmic losses so $\Qi = \Qqp$.
If the metallisation of a transmission line is superconducting, the series impedance per unit length of line, $\mathcal{Z}$, is modified to
\begin{equation}\label{eqn:tl_series_impedance}
	\mathcal{Z} = i \omega \mathcal{L}_\text{g} + g \Zs
\end{equation}
where $\mathcal{L}_\text{g}$ is the inductance per unit length in the case of PEC conductors, $\Zs = \Rs + i \Xs$ is the surface impedance of the superconductors and $g$ is a geometrical factor.
The shunt admittance per unit length is the same as the PEC case.
In general, $\Zs$ is a non-trivial function of $\sigma$.
However, for most resonators of practical interest $|\Xs| \gg \Rs$ and we may approximate
\begin{equation}\label{eqn:tl_gamma_sc}
	\gamma = \sqrt{\frac{\mathcal{Z}}{\mathcal{C}}}
	\approx \Im[\gamma] \left[ i + \frac{\kappa_\text{f} \Rs}{2\Xs} \right]
\end{equation}
where the factor $\kappa_\text{f} = g \Xs / (2 \pi i \nu \mathcal{L}_\text{g} + g \Xs)$ is normally referred to as the kinetic inductance fraction of the superconducting line.
Zmuidzinas\,\cite{zmuidzinas2012superconducting} has shown that if $\sigma_2 \gg \sigma_1$ then
\begin{equation}\label{eqn:ratio_surface_resistance_and_reactance}
	\frac{\Rs}{\Xs} \approx \kappa_\text{g} \frac{\sigma_1}{\sigma_2}
\end{equation}
where $\kappa_\text{g}$ is a scaling factor that varies in magnitude between $1/3$ and $1$ depending on the thickness of the film and whether or not it is in the extreme anomalous limit.
Combining (\ref{eqn:tl_q_from_gamma}), (\ref{eqn:tl_gamma_sc}) and (\ref{eqn:ratio_surface_resistance_and_reactance}) we again obtain the approximation $\Qqp \propto \nqp$.

\end{appendix}

\end{document}